\documentclass[10pt,a4paper]{article}
\usepackage[a4paper,top=2cm,bottom=2cm,left=1.5cm,right=1.5cm]{geometry}
\usepackage{epsfig}
\usepackage[dvipsnames]{xcolor}
\usepackage[pdftex,colorlinks,linkcolor=MidnightBlue,urlcolor=PineGreen,citecolor=Mahogany]{hyperref}
\usepackage[font={sf},labelfont={bf}]{caption}
\usepackage{helvet}
\usepackage{natbib}
\usepackage{amsmath}

\newcommand{\Bmath}[1]{ \mbox{{\boldmath ${#1}$}}  }
\newcommand{\B}[1]{\mbox{\bf #1}}
\newcommand{\msc}[1]{\textsc{#1}}

\newcommand{\grad}{\mathbf{grad}}

\newcommand{\x}{{\bf x}}

\renewcommand{\r}{{\bf r}}

\newcommand{\p}{{\bf p}}

\begin{document}

\title{3D Restoration of sedimentary terrains: The GeoChron Approach}

\author{Jean-Laurent Mallet and Anne-Laure Tertois \\
\small Emerson - Paradigm, 78 avenue du XXeme Corps, 54000 Nancy, France\\
\small anne-laure.tertois@emerson.com}


\maketitle

\date{}

\begin{abstract}
  Three-dimensional restoration of complex structural models has become
a recognized validation method. Bringing a sedimentary structural model
back in time to various deposition stages may also help understand the
geological history of a study area and follow the evolution of potential
hydrocarbon source rocks, reservoirs and closures.

Most current restoration methods rely on finite-element codes which require a mesh that
conforms to both horizons and faults, a difficult object to generate in complex
structural settings. Some innovative approaches use implicit horizon
representations to circumvent meshing requirements. In all cases, finite-element
restoration codes depend on elasticity theory which relies on mechanical
parameters to characterize rock behavior during the physical unfolding process.

In this paper, we present a geometric restoration method based on the
mathematical theory provided by the GeoChron framework. No assumption is made on
the extent of deformation, nor on the nature of terrains being restored.
Equations derived from the theory developed for the GeoChron model ensure
model consistency at each restored stage.

As the only essential input is a GeoChron model, this restoration technique does
not require any specialist knowledge and can be included in any existing
structural model-building workflow as a standard validation tool. A model can
quickly be restored to any desired stage without providing input mechanical
parameters for each layer nor defining boundary conditions, enabling geologists
to iterate on the structural model and refine their interpretations until
they are satisfied with both input and restored models.

\end{abstract}

\section{Introduction}

Many computerized methods have been developed in the past 30 years to build
numerical models of sedimentary terrains from seismic and well data, where
geological layers are often both folded and faulted. Estimations and forecasts
based on such models may impact economic decisions, so numerical
representations of available data must be as accurate and consistent as
possible.

\ \\
One way of checking whether any sub-surface model is consistent is to bring it
back in time, to a state prior to faulting and folding for a given geological
horizon \citep{Moretti2008,Maerten2015}. If such a process fails, the
incriminated areas may point out inconsistencies in the present-day structural
model. If it is successful, the geologist can use this simpler restored model
to refine his interpretations and build a geological history of the study area
\citep{DurandRiard2013}.

\ \\
Most 3D geological restoration techniques based on mechanics of continuous media
assume that geological layers deform in a linear elastic manner
\citep{Maerten2015}. However, the faulted subsurface is a discontinuous medium
in which large, non-linear plastic deformations occur. Large deformations are
taken into account by some restoration methods
(e.g.~\citet{Muron2005,Moretti2006}) but induce a time-consuming, heavy
computation load for each restored stage. Moreover, mechanical restoration
methods may result in restored models with gaps and overlaps close to
fault-induced discontinuities, which are then minimized through debatable
numerical post-processes.

\ \\
\citet{Lovely2018} present a simple, purely geometrical restoration method based
on the commercially available implementation of the GeoChron theory
\citep{Mallet2014} provided by Emerson Paradigm\textsuperscript{\textregistered} in the
SKUA\textsuperscript{\texttrademark}
software package.
In this paper, we derive a full geometrical restoration theory from the fundamental
equations of the GeoChron model and show complete implementation results.
For input GeoChron models of any degree of geometrical and topological
complexity, our method handles both small and large deformations,
does not assume elastic behavior and does not require any prior knowledge of
geo-mechanical properties. Finally, the fundamental equations of this method intrinsically integrate
minimization of gaps and overlaps along faults without any need for
post-processing.

\section{Overview of the mathematical GeoChron framework}

\begin{figure}
\centerline{\psfig{figure=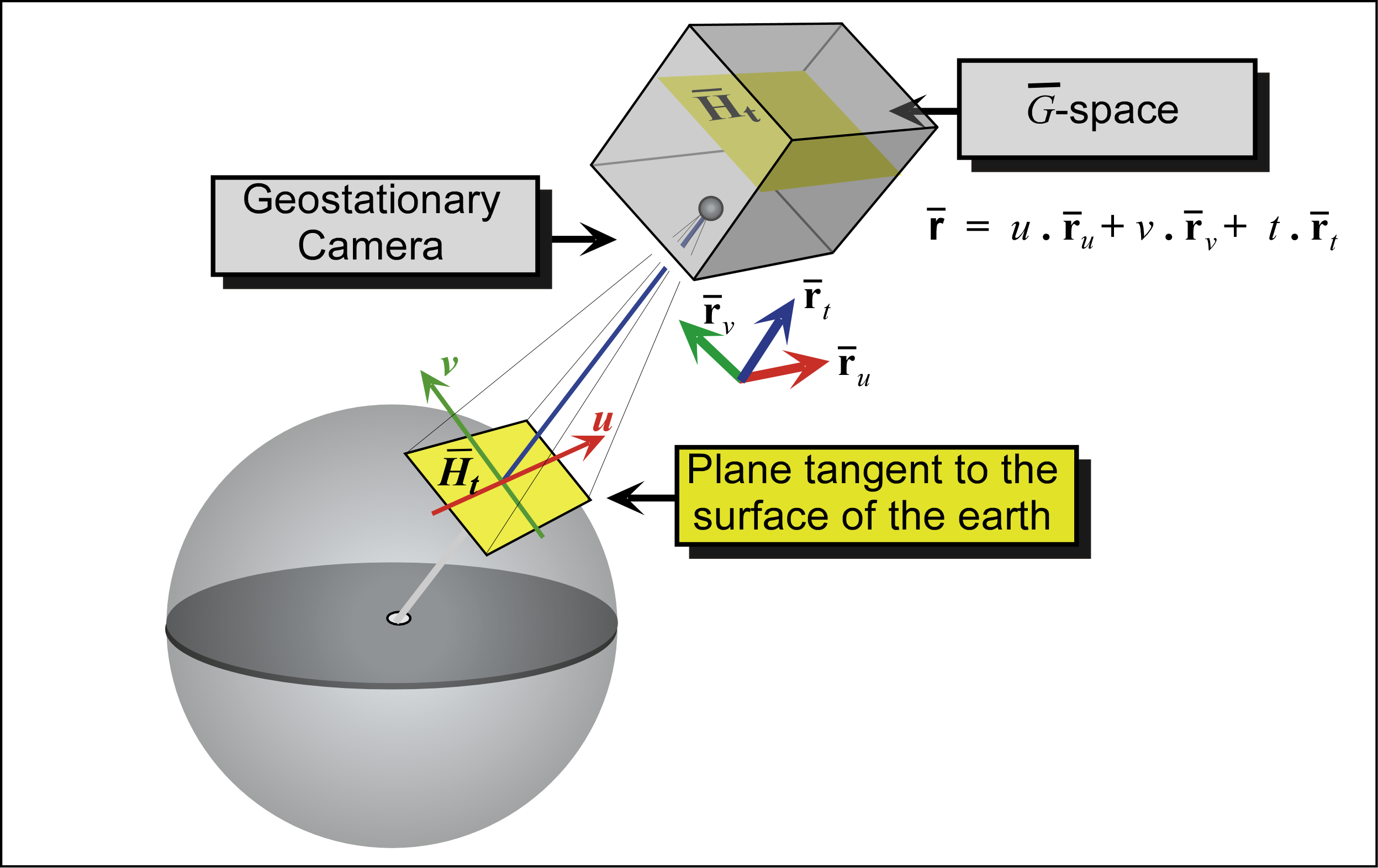,width=116mm}}
\caption{
        $\overline{G}$-space viewed as a continuous chronological stack of images $\{\overline{H}_t\}$ of
        the surface of the Earth shot from a geo-stationary camera throughout
        geological-time (in \citet{Mallet2014}, courtesy of EAGE).
        }
\label{GEOCHRONO_F2}
\end{figure}

This section briefly describes the main elements of the mathematical GeoChron
framework used in this article. The complete theory may be found in~\citet{Mallet2014}.

\ \\
As illustrated by Figure~\ref{GEOCHRONO_F2}, consider a
geo-stationary satellite pointing a camera
vertically towards a region of interest on the surface of the Earth.
This camera delimits a right-handed frame of three
orthogonal unit vectors $\{\bar\r_u, \bar\r_v, \bar\r_t\}$ where
$\bar\r_t$ is orthogonal to the surface of the Earth and oriented
upward. These three vectors define the edges of a box where
images shot by the camera are stacked in
chronological order throughout geological-time. For coherency with the
geological processes, the camera is geo-stationary in the sense
that its origin and $\{\bar\r_u,\bar\r_v,\bar\r_t\}$ vectors are ``attached''
to the tectonic plate which contains the domain of interest.
\noindent

\ \\
Let $\overline{H}_\tau$ be a horizontal plane orthogonal to the
vector $\bar\r_t$ and corresponding to the one-to-one map of the sea floor at geological-time $\tau$.
$\overline{H}_\tau$ is identical to a picture of the sea floor taken at geological-time
$\tau$ and is therefore parallel to the pair of orthogonal unit
vectors $\{\bar\r_u, \bar\r_v\}$, which can thus be used as a 2D frame
for $\overline{H}_\tau$. As a consequence, for any given
origin $\bar{\bf p}_0(t)$ belonging to $\overline{H}_\tau$,
the pair of vectors $\{\bar\r_u, \bar\r_v\}$ induces a rectilinear coordinate
system $(u,v)$ on $\overline{H}_\tau$ such that:
\begin{equation} \label{eqn:geochron:3}
\bar{\bf p}\in \overline{H}_\tau \quad \Longleftrightarrow \quad \exists\
(u,v) \in {I\!\!R}^2 \ : \
  \bar{\bf p} \ = \ \bar{\bf p}_0(\tau) + u\cdot\bar\r_u + v\cdot\bar\r_v
\end{equation}\noindent
At geological-time $\tau$, the $(u,v)$ rectilinear coordinate system
so defined can thus be used to locate on map
$\overline{H}_\tau$ any particle of sediment being deposited at that geological-time.
Therefore, the $(u,v)$ pair is called
``paleo-geographic coordinate'' system. On sea floor $H_\tau$, the reverse
image of the rectilinear coordinate axes $(u)$ and $(v)$ consists in a
curvilinear coordinate system.

\begin{figure}
\centerline{\psfig{figure=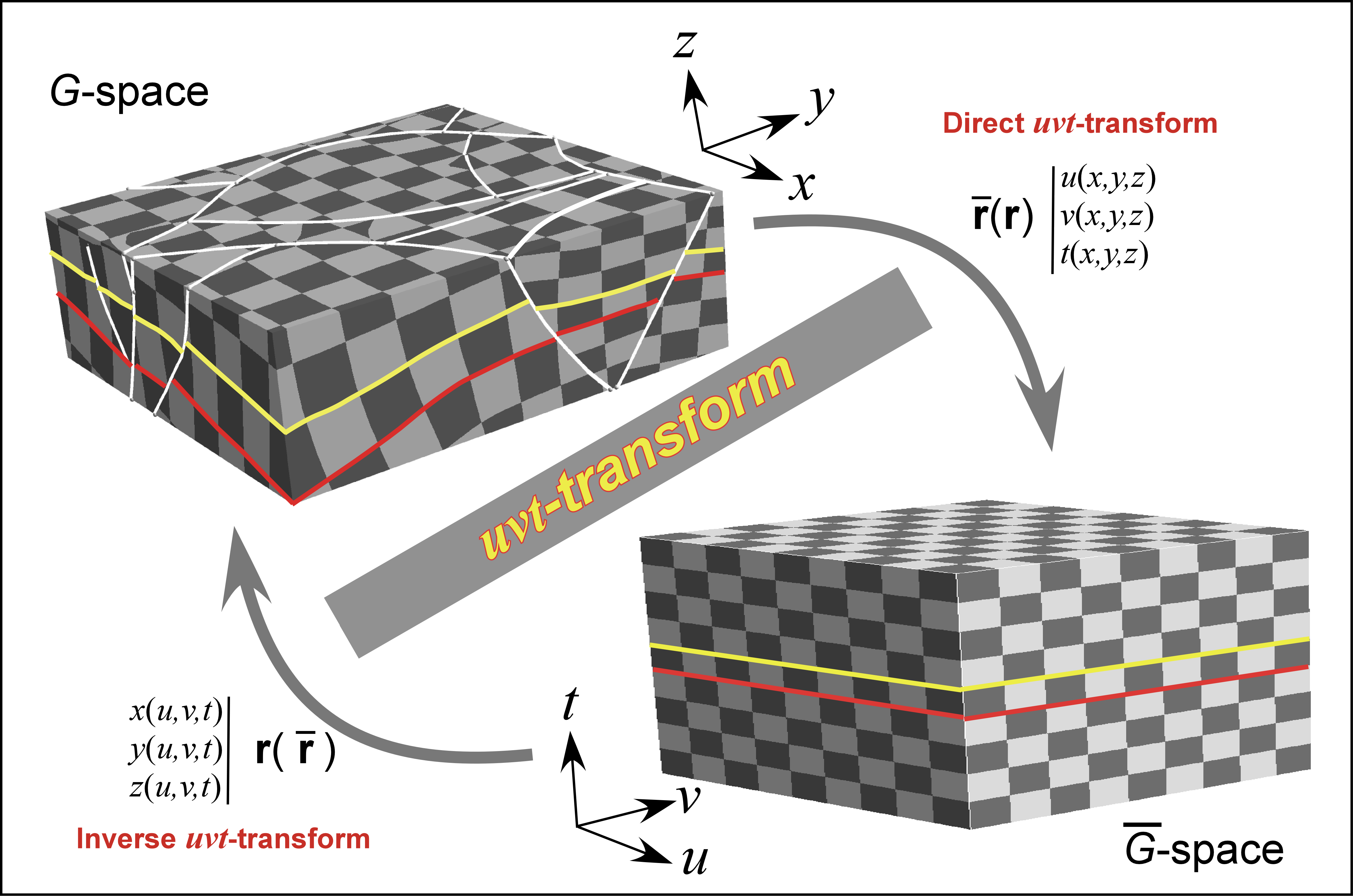,width=116mm}}
\caption{
         Graphical characterization of direct and inverse $uvt$-transforms (in
                 \citet{Mallet2014}, courtesy of EAGE).
        }
\label{GBR-UVT-transform}
\end{figure}\noindent
\ \\
Let $G$, also called $G$-space, be the domain of interest in stratified
sedimentary terrains. Two distinct coordinate systems characterize any particle
of sediment observed today at location $\r\in G$ in the subsurface:
\begin{itemize}

\item First, present-day horizontal geographic
coordinates $\{x(\r),y(\r)\}$ and altitude $z(\r)$ with respect to a given
direct 3D frame consisting of three orthogonal unit vectors $\{\r_x,\r_y,\r_z\}$
associated with the $G$-space, where $\r_z$ is vertical and oriented upward:
    \begin{equation} \label{eqn:GBR-.a1}
        \r \ = \ x(\r)\cdot\r_x + y(\r)\cdot\r_y + z(\r)\cdot\r_z
        \quad \in G
    \end{equation}\noindent

\item Second, paleo-geographic coordinates
$\{u(\r),v(\r)\}$ as they could have been observed at geological-time $t(\r)$
when the particle was deposited. These paleo-coordinates
$\{u(\r),v(\r),t(\r)\}$ define the location $\overline{\r}$ of the particle in a
``depositional'' space $\overline{G}$ also called $\overline{G}$-space:
    \begin{equation} \label{eqn:GBR-1.b}
        \overline{\r} \ = \ u(\r)\cdot\overline{\r}_u + v(\r)\cdot\overline{\r}_v + t(\r)\cdot\overline{\r}_t
        \quad \in \overline{G}
    \end{equation}\noindent
In this definition, the $\overline{G}$-space is associated with a given, direct
3D frame of three unit orthogonal vectors
$\{\overline{\r}_u,\overline{\r}_v,\overline{\r}_t\}$ where $\overline{\r}_t$ is
vertical and oriented upward. From now on, $\overline{G}$ is identified with the
box associated with the camera shown in Figure~\ref{GEOCHRONO_F2}.

\end{itemize}\noindent
As Figure~\ref{GBR-UVT-transform} shows, the equations above can be viewed as
a ``direct'' $uvt$-transform of point $\r\in G$ into a point
$\overline{\r}=\overline{\r}(\r)\in\overline{G}$ and, conversely, a ``reverse''
$uvt$-transform of point $\overline{\r}\in \overline{G}$ into a point
$\r=\r(\overline{\r})\in G$. Furthermore, the $uvt$-transform also applies as
follows to any function $\varphi$ defined in $G$:
    \begin{equation}\label{eqn:uvt-transform-phi}
    \overline{\varphi}(\overline{\r}) = \varphi(\r)\quad \forall\:\r \in G
    \end{equation}\noindent
This concept of $uvt$-transform both of points and functions plays a central
role in the GeoChron-based restoration method presented in this article. When
referring to a function $\varphi$ defined in $G$, the following
notations may be used interchangeably for clarity:
    \begin{equation} \label{eqn:geochron:dual-func-3}
    \begin{array}{llllllllllllllll}
       \quad &
       \varphi(x,y,z)
       &\equiv&
       \varphi({\r})
       &\equiv&
       \varphi\big(\ {\r}(\bar{\r}) \ \big)
    \\ \\
       \equiv&
       \bar{\varphi}(u,v,t)
       &\equiv&
       \bar{\varphi}(\bar{\r})
       &\equiv&
       \bar{\varphi}\big(\ \bar{\r}({\r}) \ \big)
       &\equiv&
       \overline{\varphi(\r)}
    \end{array}
    \end{equation}\noindent

\begin{figure}
\centerline{\psfig{figure=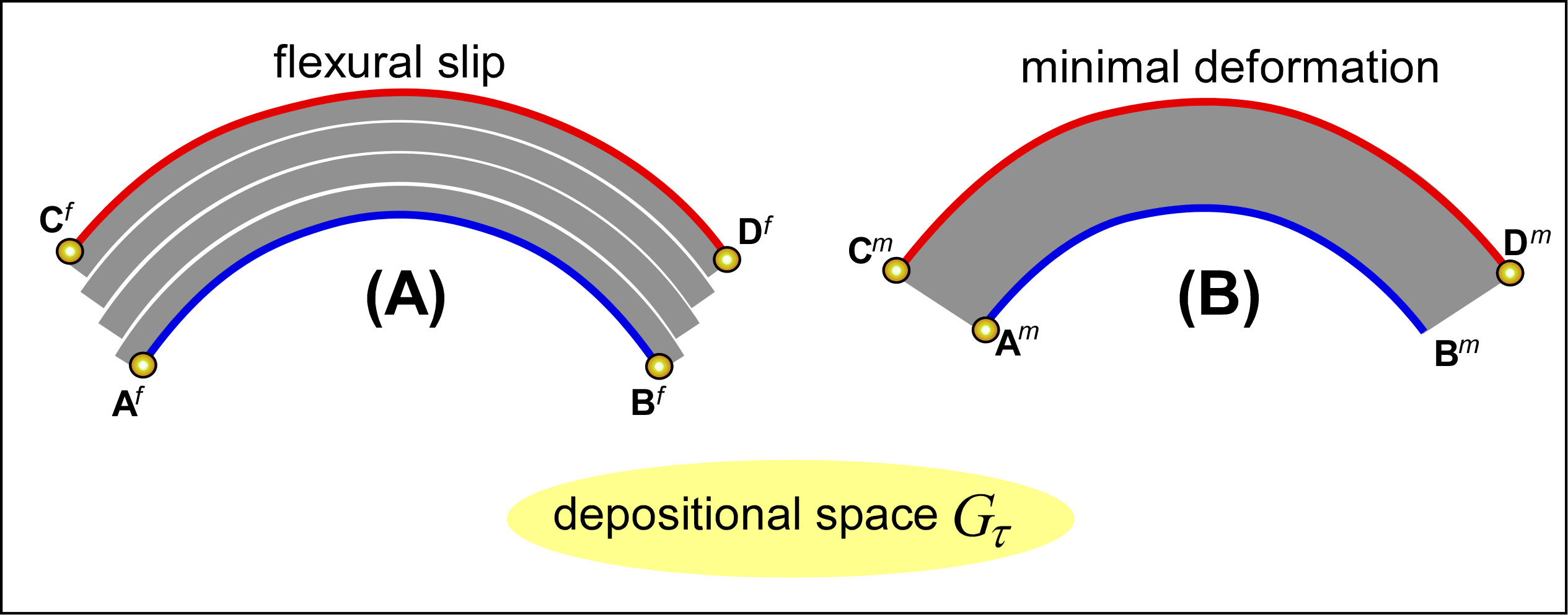,width=116mm}}
\caption{
        Vertical cross section of a cylindrical, constant thickness layer. If
        the tectonic style is flexural slip, arc lengths $A^fB^f$ and $C^fD^f$
        are equal. If the tectonic style is minimal deformation, arc length
        $A^mB^m$ is smaller than arc length $C^mD^m$. As an example, one can
        picture flexural slip occurring when the material being bent is a stack
        of paper sheets, but minimal deformation would rather occur with a piece
        of rubber.
        }
\label{GBR-Tectomic-Style}
\end{figure}\noindent
\ \\
According to geological context, a geologist can
choose one of two different tectonic styles to characterize the behavior of
geological layers subject to tectonic forces.
Figure~\ref{GBR-Tectomic-Style} illustrates both these options, referred to as
``flexural slip'' and ``minimal
deformation'' (see definitions on pages 53 and 54 of
\citet{Mallet2014}). In the GeoChron framework, these tectonic styles each
translate as a different set of equations which constrain the behavior of
paleo-geographic coordinates $\{u,v\}$.

\ \\
Let $\partial_\alpha\varphi$ denote $\partial\varphi/\partial \alpha \: \forall
\, \alpha \in \{x,y,z\}$ and $\grad\, \varphi$ denote the gradient
$\partial_x\varphi\, \r_x + \partial_y\varphi\, \r_y + \partial_z\varphi\, \r_z$
of a function $\varphi$ over frame $\{\r_x,\r_y,\r_z\}$.
The mathematical GeoChron theory presented in \citep{Mallet2014} states that,
depending on tectonic style, functions\footnote{ From now on,
we use the following concise notation: $\{f,g,\cdots\}_x\equiv\{f(x),g(x),\cdots\}$.}
$\{u,v\}_\r$
are assumed to honor, in a least squares sense, the following differential
equations (see pages~70 to 74 in \citet{Mallet2014}):
\begin{equation} \label{eqn:geochron:43E3}
    \begin{array}{cc}
    \mbox{Minimal deformation style:}
    &
    \left\{
      \begin{array}{cc}
      1 \& 2)&
      ||\grad\, u|| \simeq 1
      \quad ; \quad
      ||\grad\, v|| \simeq 1
      \\ \\
      3)&
      \grad\, u \cdot \grad\, v \ \simeq \ 0
      \\
      4)&
      \grad\, t \cdot \grad\, u \ \simeq \ 0
      \\
      5)&
      \grad\, t \cdot \grad\, v \ \simeq \ 0
      \end{array}
    \right.
    \end{array}
\end{equation}
\noindent

\begin{equation} \label{eqn:geochron:49e}
    \begin{array}{cc}
    \mbox{Flexural slip style:}
    &
    \left\{
        \begin{array}{cccccccc}
        1 \& 2)&
            ||\grad_H\,u|| \simeq 1
            \quad ; \quad
            ||\grad_H\,v|| \simeq 1
            \\ \\
        3)&
            \grad_H\,u \cdot \grad_H\,v \ \simeq \ 0
        \end{array}
    \right.
    \end{array}
\end{equation}\noindent
where $\grad\,\varphi(\r)$ is the gradient of scalar function $\varphi$ at
location $\r$,
$H\equiv H_{t(\r)}$ is the horizon passing through point
${\r\in G}$, defined as the set of particles of sediment which were deposited at
geological-time $t(\r)$,
and $\grad_H\,\varphi({\r})$ is the projection of gradient
$\grad\, \varphi({\r})$ onto said horizon $H$.

\ \\
Equations \ref{eqn:geochron:43E3} or \ref{eqn:geochron:49e} can be honored
exactly only in the particular case where horizons are perfectly planar and
parallel. In all other cases, local deformations of terrains entail that these
equations can only be approximated in a least squares sense.
As Figure~\ref{GBR-UVT-transform} shows, functions $\{u,v,t\}_\r$ are
continuous and smooth everywhere in $G$ except
across faults.

\ \\
For any equivalent system of GeoChron functions $\{u,v,t\}_\r$ and any tectonic style, it may be
shown\footnote{ See Equation~2.25 on page~64 of \citet{Mallet2014}.} that the
component ${\cal E}_{\alpha\beta}(\r)$ of the strain (deformation) tensor
$\Bmath{\cal E}(\r)$ at any point ${\r\in G}$ on the global frame
$\{\r_x,\r_y,\r_z\}$ honors the following equation:
    \begin{equation} \label{eqn:Strain-6.4}
       2\cdot {\cal E}_{\alpha\beta}(\r) \ = \
               \delta_{\alpha\beta} \ - \
               \biggl\{
                    \partial_\alpha u\cdot\partial_\beta u + \partial_\alpha v\cdot\partial_\beta v
                    + \frac{N^\alpha\cdot N^\beta}{ (1 - \phi)^2 }
               \biggr\}_{\r}
               \qquad \forall\ (\alpha,\beta)\in\{x,y,z\}^2
    \end{equation}\noindent
where $\phi({\r})$ denotes the compaction coefficient at point $\r$ defined on
page~38 of \citet{Mallet2014}
whilst $\{N^x,N^y,N^z\}_{\r}$ denote the components on $\{\r_x,\r_y,\r_z\}$ of
the unit vector $\B{N}(\r)$ orthogonal to horizon $H_{t(\r)}$ passing through
$\r$ and oriented in the direction of younger terrains:
\begin{equation} \label{GBR-N}
    \B{N}(\r) \ = \ \frac{\grad\,t(\r)}{||\grad\,t(\r)||}
\end{equation}\noindent

\section{GeoChron framework for 3D restoration}

\begin{figure}
\centerline{\psfig{figure=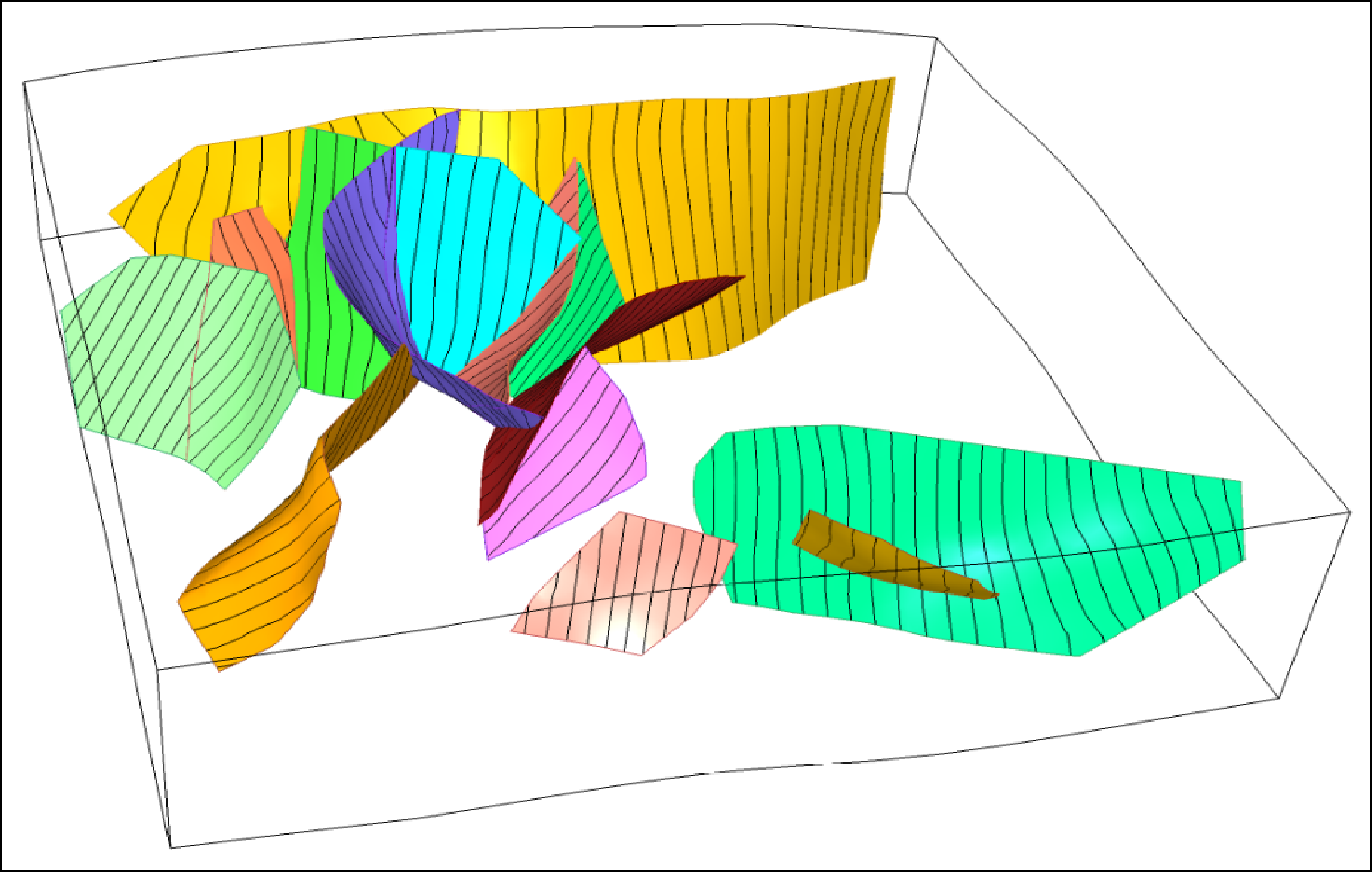,width=116mm}}
\caption{
         Example of fault-striae drawn on a fault network and deduced from
         functions $\{u,v,t\}_\r$ of a GeoChron model. Data courtesy
         of Total.
        }
\label{GBR-fault-striae}

\end{figure}

The restoration method presented in this paper uses as input an initial GeoChron
model of the studied domain $G$, which provides (see
Figure~\ref{GBR-Tetrahedral_Mesh-1} and \citet{Mallet2014}):
\begin{itemize}

\item Fault network topology and geometry;

\begin{figure}
\centerline{\psfig{figure=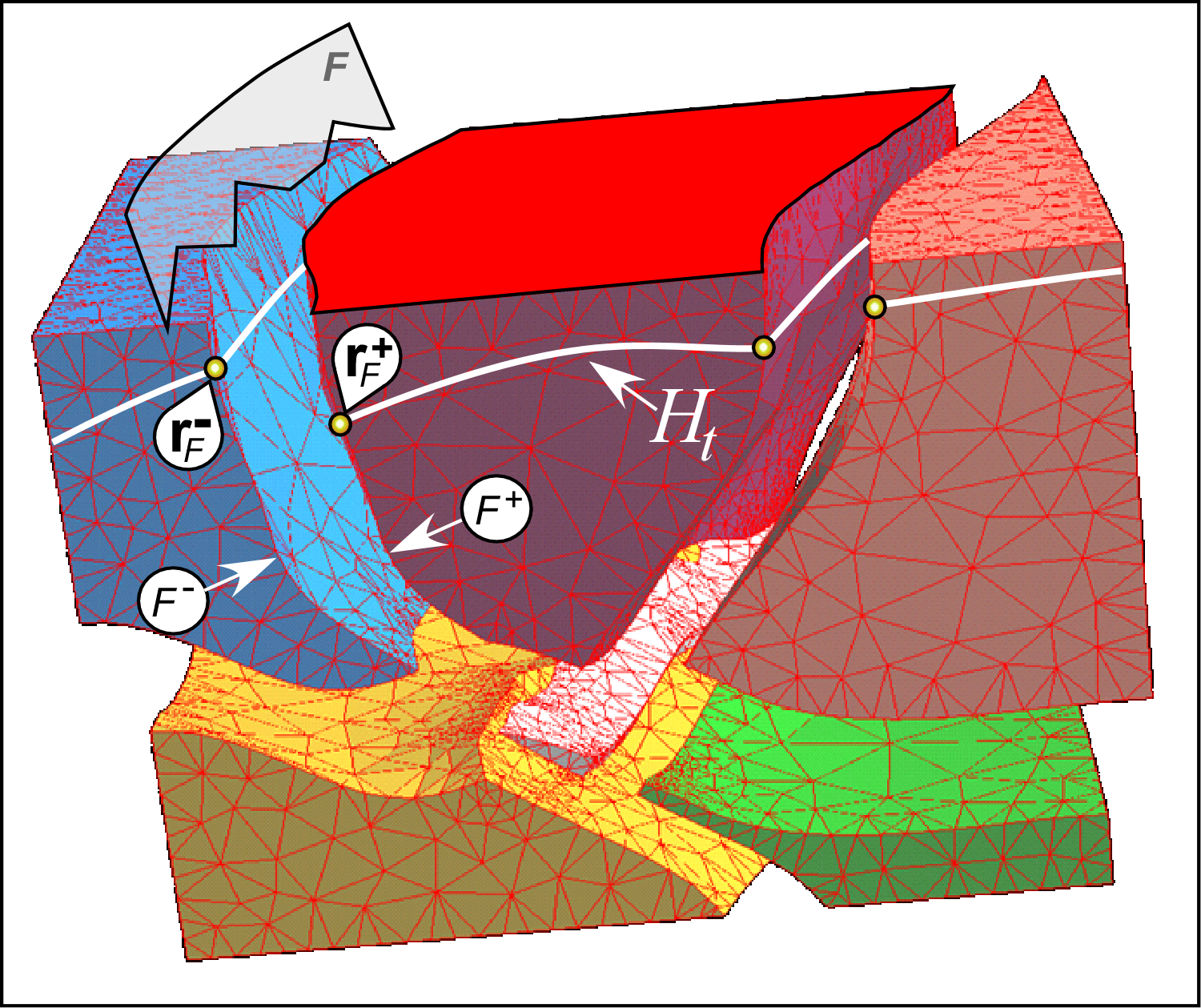,width=116mm}}
\caption{
         Exploded view of a faulted, 3D geological domain $G$.
         During restoration, twin faces $F^-$ and $F^+$ of fault $F$
         must slide on one another. Points $(\r_{\msc{f}}^+,\r_{\msc{f}}^-)$, which
         were collocated on horizon $H_t$ at deposition time $t$ prior to
         faulting, are denoted a pair of ``twin-points''.
        }
\label{GBR-Tetrahedral_Mesh-1}
\end{figure}\noindent

\item For each geological fault $F$, two disconnected surfaces $F^+$ and
$F^-$ bordering $F$ on either side. As observed today, $F^+$ and $F^-$ are collocated; however,
during the restoration process, $F^+$ and $F^-$ should slide on one another,
without generating gaps or overlaps between adjacent fault blocks;

\item For each fault $F$, a set of pairs of points $(\r_{\msc{f}}^+,\r_{\msc{f}}^-)$
called ``twin-points'' and such that:
    \begin{enumerate}
        \item $\r_{\msc{f}}^+\in F^+$ and $\r_{\msc{f}}^-\in F^-$;
        \item Before $F$ induced any movement in the subsurface, the particles
        of sediment which are observed today at locations $\r_{\msc{f}}^+$ and
        $\r_{\msc{f}}^-$ were collocated.
    \end{enumerate}
    \noindent
    During the activation of fault $F$, particles of sediment initially located
    on $F$ are assumed to slide along apparent fault-striae defined as the
    shortest path, on $F$, between pairs of twin points
    $(\r_{\msc{f}}^+,\r_{\msc{f}}^-)$ (see example of fault network with
    fault-striae in Figure~\ref{GBR-fault-striae}). From now on and for concision's sake,
    ``apparent'' fault-striae will simply be called ``fault-striae'';

\item A tectonic style which may be either ``minimal deformation'' or
``flexural slip'';

\item A triplet $\{u,v,t\}_\r$ of piecewise continuous functions defined on
the $G$-space such
that, for a particle of sediment observed today at location $\r\in G$, the numerical
values $\{u(\r),v(\r)\}$ represent the paleo-geographic coordinates of the particle
at geological-time $t(\r)$ when it was deposited.

\end{itemize}\noindent
Moreover, inherent to the GeoChron model, the following points are of relevance
to the restoration method presented in this paper:
\begin{itemize}
\item Each geological horizon $H_{\tau}$ is the set of particles of
sediment deposited at a given geological-time $\tau$:
        \begin{equation} \label{eqn:Restoration-00}
            \r\in H_{\tau}
            \quad \Longleftrightarrow \quad
            t(\r) = \tau
        \end{equation}
        \noindent
        In other words, $H_{\tau}$ is defined as a level-set of the
        geological-time function $t(\r)$;

\item Paleo-geographic coordinates functions $\{u,v\}_\r$ and twin-points are
linked by the following equations:
        \begin{equation} \label{GBR:TP-1}
            \begin{array}{c}
                    \{\ (\r_{\msc{f}}^+,\r_{\msc{f}}^-) \mbox{ is a pair of twin-points } \}
                    \quad \Longleftrightarrow \quad
                    \left\{
                    \begin{array}{ccccccc}
                        1)& \r_{\msc{f}}^+ \in F^+ \ \mbox{ \B{\&} } \ \r_{\msc{f}}^- \in F^- \\ \\
                        2)&  u(\r_{\msc{f}}^-) = u(\r_{\msc{f}}^+)  \\
                        3)&  v(\r_{\msc{f}}^-) = v(\r_{\msc{f}}^+)  \\
                        4)&  t(\r_{\msc{f}}^-) = t(\r_{\msc{f}}^+)
                    \end{array}
                    \right.
            \end{array}
        \end{equation}\noindent
As shown in Figure~\ref{GBR-fault-striae}, each pair of twin-points $(\r_{\msc{f}}^+,\r_{\msc{f}}^-)$
is the intersection of a level set of function $t(\r)$ with a
fault-stria. As a consequence of constraints 2-3-4 above, fault-striae
characterize the paleo-geographic coordinates $\{u,v\}_\r$, and vice versa;

\item Each point $\r\in G$ is characterized by its coordinates
$\{x(\r),y(\r),z(\r)\}$ with respect to a direct frame $\{\r_x,\r_y,\r_z\}$ consisting
of three mutually orthogonal unit vectors with $\r_z$ oriented upward;

\item At any location $\r$ in geological domain $G$, the
equation $\x(s|\r)$ of curve ${\cal N}(\r)$ passing through $\r$, denoted ``normal-line'',
where $s$ is the arc length abscissa along ${\cal N}(\r)$, obeys the following
differential relationship:
\begin{equation} \label{GBR:N}
   \frac{d \x(s|\r)}{ds} \ = \ \B{N}(\x(s|\r))
\end{equation}\noindent
where $\B{N}(\r)$ is the unit vector defined by equation \ref{GBR-N}.

\end{itemize}

\subsection*{Problem to address}

\label{Restoration-PB}

Let us assume that, at given geological-time $\tau$, horizon $H_{\tau}$ to be
restored coincided with a given, smooth surface $\widetilde{H}_\tau$ considered
as the sea floor, whose altitude at geological-time $\tau$ is a given
function\footnote{ In practice, $z_\tau^o(u,v)$ should be negative everywhere in the studied domain.}
$z_\tau^o(u,v)$ of GeoChron paleo-geographic coordinates $(u,v)$:
    \begin{equation} \label{eqn:Restoration-So}
        \{ \ \r_\tau^o\in{H}_\tau \ \}
        \quad \Longleftrightarrow \quad
        \{ \ z_\tau^o(\r_\tau^o) = z_\tau^o(u(\r_\tau^o),v(\r_\tau^o)) \ \}
    \end{equation}\noindent
Let $G_\tau$ be the part of the $G$-space observed today and geologically
deposited up to geological-time $\tau$:
    \begin{equation} \label{eqn:Restoration-Gtau}
    \begin{array}{|c|} \hline \\
    \quad
          \r_\tau\in G_\tau \quad \Longleftrightarrow \quad t(\r_\tau)\leq \tau
    \quad
    \\ \\ \hline \end{array}
    \end{equation}\noindent
The problem then consists in:

\begin{enumerate}

\item Restoring horizon $H_{\tau}$ to its initial, unfolded and unfaulted state
$\widetilde{H}_\tau$ corresponding to sea floor $\widetilde{\cal S}_\tau(0)$ at geological-time $\tau$;

\item Reshaping the terrains in such a way that, for each point $\r_\tau\in G_\tau$ stratigraphically located below $H_\tau$:
    \begin{enumerate}

    \item the particle of sediment currently located at point $\r_\tau$ moves to its
    former, restored location $\{\bar\r_\tau=\bar\r_\tau(\r_\tau)\}$, where it was at geological-time
    $\tau$;

    \item no overlaps or voids are created in the subsurface;

    \item volume variations are minimized whilst also taking compaction into account.

    \end{enumerate}

\end{enumerate}

\begin{figure}
\centerline{\psfig{figure=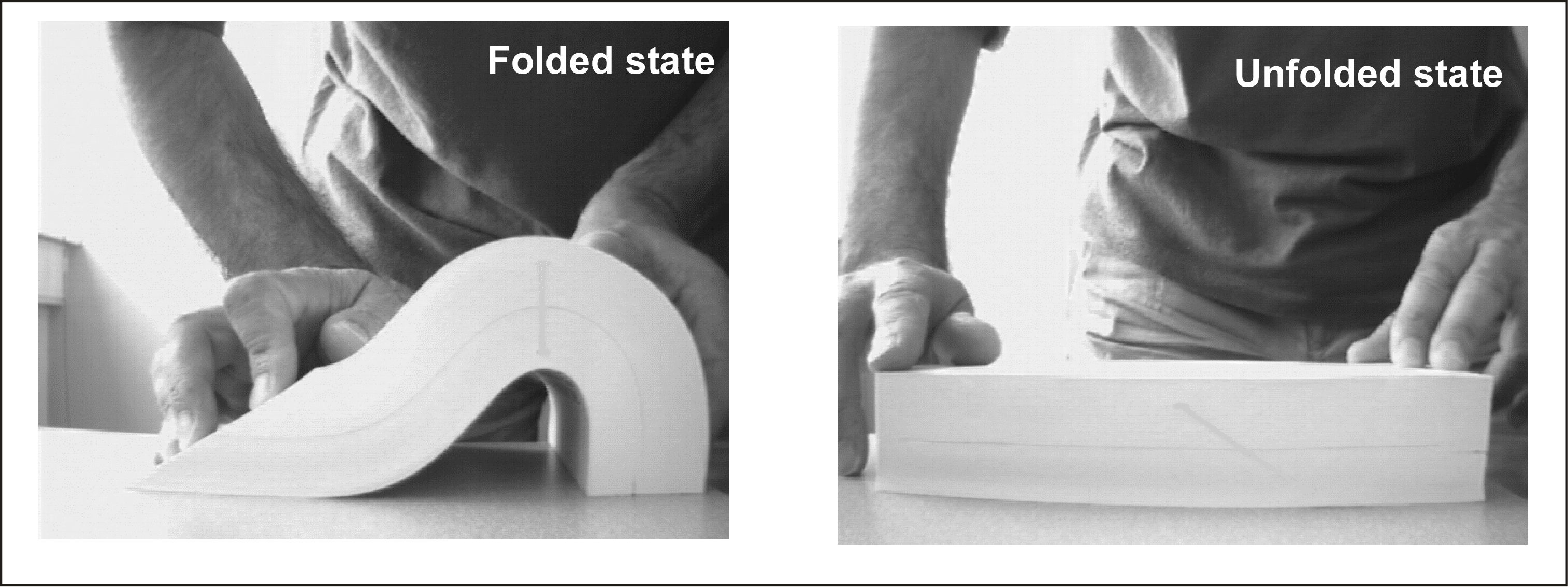,width=116mm}}
\caption{
        Folding / unfolding a book. Modified after \citet{Mallet2014}.
        }
\label{GBRestoration-XX}
\end{figure}

\subsection*{Comment}

Figure~\ref{GBRestoration-XX} shows a book, considered as the analogue of a
stack of geological layers, being folded, then laid flat again. Two distinct
types of differential equations drive the book's geometrical transformations:

\begin{itemize}

\item \B{Folding:}
    \begin{enumerate}
        \item Initial state:
            \begin{itemize}
                \item Book pages are parallel and flat, which implies that the
                book's initial geometry is known;
                \item Mechanical laws which control the behavior of the pages
                (e.g., elasticity) are known;
                \item Physical properties (e.g., Lam\'e coefficients) of the
                pages are known;
                \item External forces applied to the book are known.
            \end{itemize}\noindent
        \item Final state:
            \begin{itemize}
                \item The book folds under the action of given external forces;
                \item Its final geometry may be deduced from the aforementioned
                mechanical laws and physical properties.
            \end{itemize}\noindent
    \end{enumerate}\noindent
    In this first case, the geometry of the final state is dictated by a set of
    differential equations controlled by initial geometry and mechanical
    properties. It seems quite obvious that, with the application of similar
    external forces, the book's final geometry will differ considerably
    according to whether its pages are made of paper, plastic or steel.

\item \B{Unfolding:}
    \begin{enumerate}
        \item Initial state:
            \begin{itemize}
                \item Book pages are folded and their geometry is given;
            \end{itemize}\noindent
        \item Final state:
            \begin{itemize}
                \item The top page of the book is flat and its geometry is given,
                \item All book pages remain parallel, without any void or intersection.
            \end{itemize}\noindent
    \end{enumerate}\noindent
    In this second case, the geometry of the final state is dictated by a set of
    differential equations controlled by initial and final conditions only and
    does not depend on the pages' mechanical properties.

\end{itemize}
\noindent
This analysis shows that differential equations which rule the folding and
unfolding cases differ and do not require the same input and boundary
conditions. In particular, unfolding does not require the mechanical properties
of the medium (pages of the book) to be known. This is why we state that
geologic restoration can be purely geometrical, without relying on
geo-mechanical laws and physical properties of geologic layers.

\subsection*{Prior art}

Since seminal article \citet{Dahlstrom1969} was published half a century ago,
dozens of methods have been proposed to restore sedimentary terrains as they
were at a given geologic time $\tau$ (e.g. \citet{Gibbs1983, Suppe1985, Muron2005,
Moretti2006, Moretti2008, Maerten2015}), including some that represent
horizons as level-sets of a geological-time function
\citep{DurandRiard2010}. So far, only the one developed by \citet{Lovely2018} is
based on the GeoChron model paradigm.

\ \\
Figure~\ref{GBR-UVT-transform} illustrates that the $uvt$-transform of the
subsurface has the general look of restored stratified terrains. However, this
is not true restoration because, in the ``unfolded'' $\overline{G}$-space, all
horizons are transformed into parallel, horizontal planes, so lateral variations
in layer thicknesses are generally not preserved.

\ \\
Note that, barring compaction, in the very particular case where all layers have
a constant thickness and the following equation holds
    \begin{equation} \label{GBR:THGrad}
       ||\grad\,t_\tau(\r_\tau)|| = 1
       \qquad \forall\ \r_\tau\in G_\tau
    \end{equation}\noindent
then, the $uvt_\tau$-transform $\overline{G}_\tau$ of $G_\tau$ preserves the
thickness of each layer, which implies that $\overline{G}_\tau$ so obtained
could be considered as a restored version of $G_\tau$. This observation led to
the following comment on page~91 in \citet{Mallet2014}, recalled here:


\[
\left.
\begin{array}{c}
    \mbox{
        \begin{minipage}[t]{115mm}
            \begin{small}\sl
            $\ll$
                [\ldots] using a GeoChron model as an \B{input}, \\ it is possible to develop new breeds of unfolding algorithms.
            $\gg$
            \end{small}
        \end{minipage}\noindent
        }
\end{array}
\right.
\]
\noindent
By adding minimal functionalities to commercial
SKUA\textsuperscript{\tiny\textregistered} software designed to implement the
GeoChron model, \citet{Lovely2018} proposed a first, easy to implement
restoration algorithm which consists in using classical GeoChron equations to
compute restoration functions.
These ``native'' GeoChron equations were not devised with restoration of
sedimentary terrains as a goal. Despite this, by clever use of the software,
\citeauthor{Lovely2018} achieved remarkable first restoration results. As a
remedy to some weaknesses their study pointed out, in this paper we adapt the
GeoChron model theory, rather than its implementation.
The new set of differential equations and boundary conditions
obtained as a result are specifically designed to solve geometric restoration
problems and provide geologically and geometrically consistent
restored models. As we develop each step in our method, we will point out the
differences with \citeauthor{Lovely2018}'s work.

\section{GeoChron-Based Restoration (GBR) }

\label{GBR-GBR}

\begin{figure}
\centerline{\psfig{figure=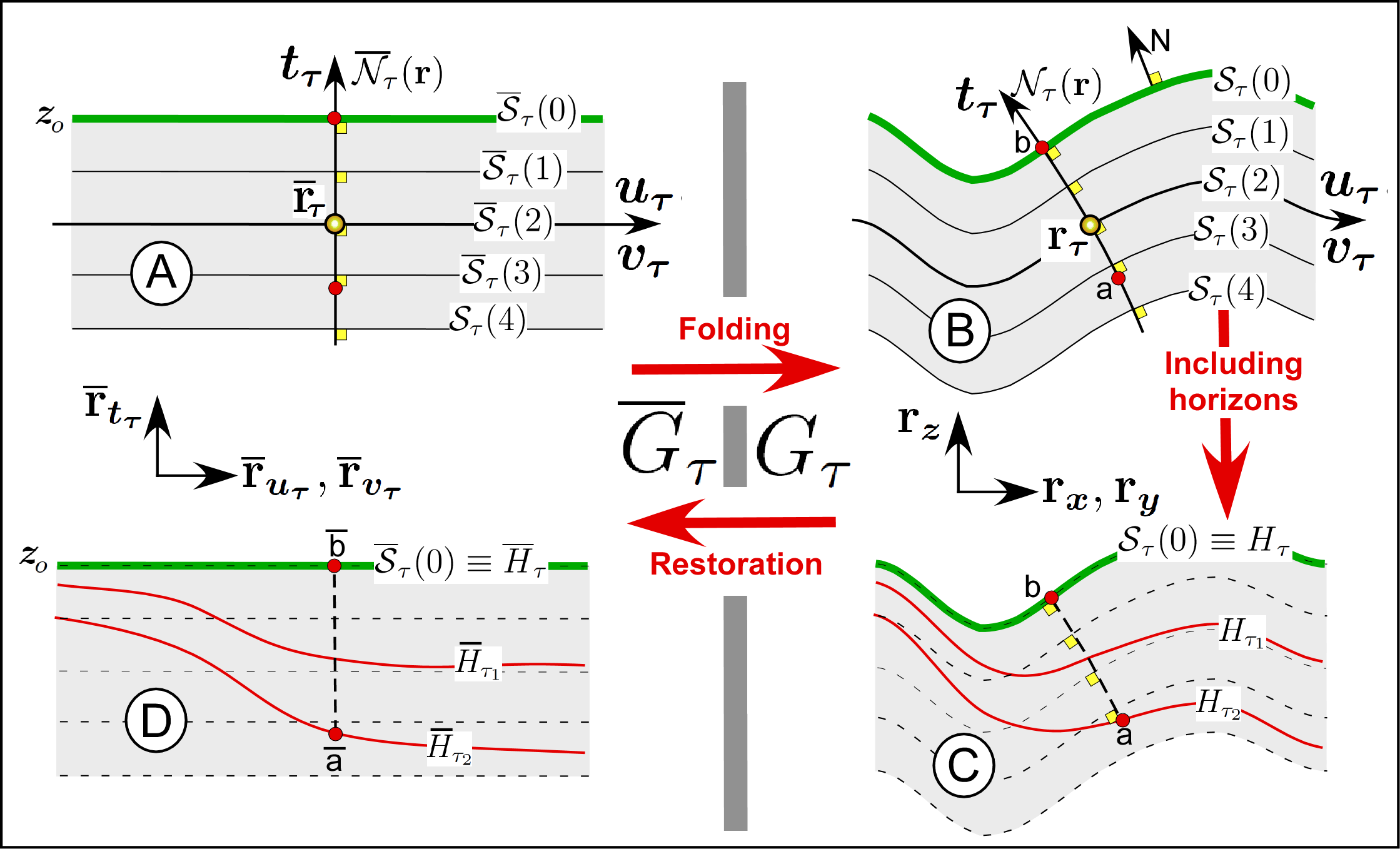,width=140mm}}
\caption{
        Jelly principle (minimal deformation style):
        The $u_\tau v_\tau t_\tau$-transform of jelly block $G_\tau$
        consists of a jelly block $\overline{G}_\tau$ identical to the restored
        version of $G_\tau$ as it was at geological-time $\tau$.
        In \citet{Tertois2019}.
        }
\label{GBRestoration-1}
\end{figure}

This section describes a purely geometrical method directly derived from the
GeoChron mathematical framework and aimed at restoring terrains at a given
geological-time $\tau$, whatever the structural complexity of horizons and
faults in studied domain $G$. This GeoChron-Based Restoration (GBR) method can be
intuitively introduced by the ``jelly block'' analogy depicted in
Figure~\ref{GBRestoration-1}.

\subsection*{Jelly block $\Bmath{\overline{G}_\tau}$}

Figure~\ref{GBRestoration-1}-A shows an arbitrarily-shaped ``jelly block'',
denoted $\overline{G}_\tau$, which contains a direct frame of orthogonal unit
vectors $\{\bar\r_{u_\tau},\bar\r_{v_\tau},\bar\r_{t_\tau}\}$
and a family of smooth, continuous horizontal surfaces $\{\overline{\cal S}_\tau(d):d\geq 0\}$
intersected only once by any vertical straight line parallel to $\bar\r_{t_\tau}$:

\begin{enumerate}

\item For each point ${\bar\r_\tau}\in \overline{G}_\tau$,
$\{u_\tau({\bar\r_\tau}),v_\tau({\bar\r_\tau})\}$ represent the horizontal
coordinates of ${\bar\r_\tau}$ with respect to horizontal unit frame vectors $\{\bar\r_{u_\tau},
\bar\r_{v_\tau}\}$ whilst $t_\tau(\bar\r_\tau)$ represents its altitude with respect
to $\bar\r_{t_\tau}$, oriented upward;

\item Any point $\bar\r_\tau^o\in\overline{\cal S}_\tau(0)$ is located at altitude
$(t_\tau(\bar\r_\tau^o)=0)$ with respect to the vertical unit vector $\bar\r_{t_\tau}$
oriented upward;

\item $\overline{\cal S}_\tau(d)$ is located at algebraic vertical
distance $(d)$ from $\overline{\cal S}_\tau(0)$ in such a way that:
                \begin{equation} \label{GBR:X0001}
                   \left|
                   \begin{array}{cccccccc}
                        d<0 &\Longleftrightarrow& \overline{\cal S}_\tau(d) & \mbox{is located above } \overline{\cal S}_\tau(0)
                        \\
                        d>0 &\Longleftrightarrow& \overline{\cal S}_\tau(d) &
\mbox{is located below } \overline{\cal S}_\tau(0)
                   \end{array}
                   \right.
                \end{equation}\noindent

\end{enumerate}
\noindent
Contrarily to classical Free Form Deformation methods which, following the
principles formulated by \citet{Sederberg1986}, introduce the
concept of jelly block, $\overline{G}_\tau$ may be of arbitrary shape and may be
discontinuous across surfaces dividing it, either partially or totally.

\subsection*{Jelly block $\Bmath{G_\tau}$}

Figure~\ref{GBRestoration-1}-B shows the
folded jelly block $G_\tau$ resulting from the deformation of jelly block
$\overline{G}_\tau$ under tectonic forces induced either by minimal deformation
or flexural slip tectonic forces:
                \begin{equation} \label{GBR:X2}
                \begin{array}{|c|} \hline \\
                \quad
                    \begin{array}{c}
                       \overline{G}_\tau \quad \longrightarrow\mbox{Tectonic
                           forces}\longrightarrow \quad G_\tau
                    \end{array}
                \quad
                \\ \\ \hline \end{array}
                \end{equation}\noindent
In spaces $\overline{G}_\tau$ and $G_\tau$:

\begin{enumerate}

\item Using reverse and direct $u_\tau v_\tau t_\tau$-transforms,
each point ${\bar\r_\tau}\in\overline{G}_\tau$ is transformed into point
$\r_\tau\in G_\tau$, and conversely:
                \begin{equation} \label{GBR:X1a}
                  {\bar\r_\tau}\in\overline{G}_\tau \quad \longleftrightarrow
                  \quad \r_\tau\in G_\tau
                \end{equation}\noindent

\item For each point $\r_\tau\in G_\tau$:
        \begin{itemize}

        \item $\{x(\r_\tau),y(\r_\tau)\}$ represent the horizontal geographic coordinates
        of $\r_\tau$ with respect to $\{\r_x,\r_y\}$, whilst $z(\r_\tau)$ represents its
        altitude with respect to the vertical unit frame vector $\r_z$ oriented
        upward;

        \item $\{u_\tau,v_\tau,t_\tau\}_{\r_\tau}$ are functions defined as follows in $G_\tau$:
        \begin{equation} \label{GBR-XXX-1}
          \begin{array}{|c|} \hline \\
           u_\tau(\r_\tau) = u_\tau(\bar{\r}_\tau) \quad ; \quad
           v_\tau(\r_\tau) = v_\tau(\bar{\r}_\tau) \quad ; \quad
           t_\tau(\r_\tau) = t_\tau(\bar{\r}_\tau)
          \\ \\ \hline
          \end{array}
        \end{equation}
        \end{itemize}\noindent

\item Each horizontal surface $\overline{\cal S}_\tau(d)\in \overline{G}_\tau$ is
transformed into a curved surface ${\cal S}_\tau(d)\in G_\tau$ ``parallel\footnote{ The
notion of ``parallelism'' is linked to eikonal Equation~\ref{GBR:eikonal-equation}.}'' to ${\cal S}_\tau(0)$ and each surface ${\cal S}_\tau(d)$ is a level-set of function $t_\tau(\r_\tau)$;

\item The images of rectilinear coordinate axes $(u_\tau)$, $(v_\tau)$ and
$(t_\tau)$ contained in jelly block $\overline{G}_\tau$ consist of curved lines
in folded jelly block $G_\tau$.

\end{enumerate}
\noindent
From now on, without loss of generality and for the sake of simplicity,
$\overline{G}_\tau$-space frame $\{\bar\r_{u_\tau},\bar\r_{v_\tau},\bar\r_{t_\tau}\}$
and its origin
$\overline{O}_{u_\tau v_\tau t_\tau}$ are identified with $G$-space frame
$\{\r_x,\r_y,\r_z\}$ and its origin $O_{xyz}$:
                \begin{equation} \label{GBR:X00}
                   \bar\r_{u_\tau} \equiv \r_x \quad;\quad
                   \bar\r_{v_\tau} \equiv \r_y \quad;\quad
                  \bar\r_{t_\tau} \equiv \r_z \quad;\quad
                   \overline{O}_{u_\tau v_\tau t_\tau} \equiv O_{xyz}
                \end{equation}\noindent
Equivalently to Equations~\ref{GBR:X00}, we can state that the jelly particle
observed at location $\r_\tau\in G_\tau$ may be moved (i.e. restored) to its
former, initial location $\bar\r_\tau=\bar\r_\tau(\r_\tau)$
defined as follows, where $\B{R}_\tau(\r_\tau)$ is called ``restoration vector
field'':
                \begin{equation} \label{GBR:X00.a}
                \begin{array}{|c|} \hline \\
                    \begin{array}{c}
                        \bar\r_\tau(\r_\tau) \ = \ \r_\tau \ + \ \B{R}_\tau(\r_\tau)
                        \qquad \forall\ \r_\tau\in G_\tau
                        \\ \\
                        \mbox{with : }\quad
                        \B{R}_\tau(\r_\tau) = [\r_x, \r_y, \r_z] \cdot
                        \left[ \hspace{-2mm}
                        \begin{array}{c}
                            u_\tau(\r_\tau)-x(\r_\tau)   \\
                            v_\tau(\r_\tau)-y(\r_\tau)   \\
                            t_\tau(\r_\tau)-z(\r_\tau)
                        \end{array}
                        \hspace{-2mm} \right]
                    \end{array}
                \\ \\ \hline \end{array}
                \end{equation}\noindent


\subsection*{Fundamental GeoChron-Based Restoration principle}

We can conclude from the statements above that jelly block $G_\tau$ may be
considered as a pseudo-subsurface whose geometry at time of deposition $\tau$
was identical to $\overline{G}_\tau$ and where all pseudo-horizons $\{{\cal S}_\tau(d): d\geq 0\}$ are assumed
to be parallel. Therefore, for any point $\r$ within jelly
block $G_\tau$, restoration functions $u_\tau(\r_\tau)$ and $v_\tau(\r_\tau)$ may be identified with pseudo
paleo-geographic coordinates and $t_\tau(\r_\tau)$ may be identified with a pseudo
geological-time of deposition, which leads us to derive the following ``fundamental GeoChron-based restoration principle'':
\begin{equation}\label{GBR-fundamental-principle}
\begin{array}{|c|} \hline
    \mbox{
        \begin{minipage}[t]{130mm}
            \ \\
            {\sc Fundamental GBR principle: } \\ \sl
                Barring the effects of compaction, equations established for the
                GeoChron functions $\{u,v,t\}_\r$ also apply to functions $\{u_\tau,v_\tau,t_\tau\}_{\r_\tau}$
        \end{minipage}\noindent
        }
\\ \\ \hline \end{array}
\end{equation}

\subsection*{GeoChron-Based Restoration (GBR) algorithm }

Assume that a numerical GeoChron model characterized by functions $\{u,v,t\}_\r$
defined on a possibly faulted geological domain $G$ is given. To restore the terrains to their
state at given restoration geological-time $\tau$, as Figure~\ref{GBRestoration-1} shows, the following GeoChron-Based Restoration algorithm is proposed:
\begin{enumerate}

\item Identify the part of the subsurface stratigraphically located below
horizon $H_\tau$ with jelly block $G_\tau$.

\item Identify our restoration problem with a jelly block restoration problem.
For that purpose, make the following assumptions:
    \begin{enumerate}

        \item reverse $u_\tau v_\tau t_\tau$-transform ${S}_\tau(0)$ of 
        sea floor $\overline{S}_\tau(0)$ is identified with horizon $H_\tau$:
                \begin{equation} \label{GBR-H-tau}
                      {\cal S}_\tau(0) \ \equiv \ H_\tau
                \end{equation}

        \item at depositional time $\tau$, $\overline{S}_\tau(0)$ is
        temporarily assumed to be flat and horizontal and is identified with sea
        level with an altitude of zero; in other words, the following temporary
        assumption is made:
                \begin{equation} \label{GBR-G-tau-A}
                      t_\tau(\r_\tau^o) \ = \ 0
                      \qquad \forall\ \r_\tau^o \in H_\tau
                \end{equation}
                \noindent

        \item in $G_\tau$, terrain compaction is temporarily ignored;
    \end{enumerate}\noindent

\item Using the jelly block paradigm, to restore subsurface geometry to
geological-time $\tau$:
    \begin{enumerate}

        \item using equations and numerical techniques described in sections
        \ref{BGR:Characterizing-t-tau} to \ref{GBR-taking-faults-into-account},
        compute numerical approximations of functions
        $\{u_\tau,v_\tau,t_\tau\}_{\r_\tau}$ on $G_\tau$;

        \item compute restoration vector field $\B{R}_\tau(\r_\tau)$ defined by
        Equation~\ref{GBR:X00.a} and generate restored jelly block
        $\overline{G}_\tau$ as the $u_\tau v_\tau t_\tau$-transform of $G_\tau$;
                \begin{equation} \label{GBR-R-tau}
                      \bar\r_\tau(\r_\tau)\in \overline{G}_\tau  \quad
\Longleftrightarrow \quad \bar\r_\tau(\r_\tau) = \r_\tau + \B{R}_\tau(\r_\tau)
                \end{equation}

        \item using a specific algorithm described in section
        \ref{GBR:::compaction}, reverse compaction assumption \#2.c;

        \item to reverse the flat $\overline{S}_\tau(0)$ assumption \#2.b, move each point $\overline{\r}_\tau\in
        \overline{G}_\tau$ downward\footnote{ The sea floor is located below the
        sea level which implies that $z_\tau^o$ is constantly negative.} as follows
                \begin{equation} \label{GBR-T-tau}
                      \overline{\r}_\tau \quad \longleftarrow \quad
                      \overline{\r}_\tau \ + \
                      \{
                            t_\tau(\overline{\r}_\tau) + z_\tau^o( u_\tau(\overline{\r}_\tau),v_\tau(\overline{\r}_\tau))
                      \} \cdot \overline{\r}_{t_\tau}
                      \qquad \forall\ \overline{\r}_\tau\in \overline{G}_\tau
                \end{equation}
            where $z_\tau^o(u,v)$ is assumed to be a given function of GeoChron
            paleo-geographic coordinates; In practice, $z_\tau^o(u,v)$ may be
            defined on $H_\tau$.

    \end{enumerate}\noindent

\end{enumerate}
\noindent
At first glance, replacing $(u,v)$ by $(u_\tau,v_\tau)$ in
Equation~\ref{GBR-T-tau} may seem dubious. To justify this, in
$\overline{G}_\tau$, consider the vertical straight line
$\overline{\Delta}(u_\tau,v_\tau)$ with constant paleo-geographic coordinates
$(u_\tau,v_\tau)$. The straight line $\overline{\Delta}(u_\tau,v_\tau)$ so
defined cuts the horizontal plane $\overline{\cal S}_\tau(0)$ at a point with
paleo-geographic coordinates $(u_\tau,v_\tau,t_\tau=0)$. The crux point of our
argument is that, if the restoration process is coherent with the input GeoChron
model then, on $\{\overline{S}_\tau(0)\equiv\overline{H}_\tau\}$,
paleo-geographic coordinates $(u_\tau,v_\tau)$ are exactly the
same\footnote{ See Equations~\ref{GBR:Indet}.} as the GeoChron paleo-geographic
coordinates $(u,v)$. Therefore, Equation~\ref{GBR-T-tau} simply states that, in
$\overline{G}_\tau$, the entire column of sediments located on line
$\overline{\Delta}(u_\tau,v_\tau)$ is rigidly moved downward in such a way that
the particle of sediment at the top of this column, which was at altitude zero of
sea level, is moved to the correct, given altitude of the sea floor at
geological-time $\tau$.

\subsection*{Preservation of GeoChron functions}

In the proposed GBR process, the ``true'' GeoChron paleo-geographic coordinate
functions $\{u,v\}_\r$ and ``true'' geological-time function $t(\r)$ of the
GeoChron model provided as input are transformed passively. In other words,
after restoration, paleo-geographic coordinates
$\{u(\r_\tau),v(\r_\tau),t(\r_\tau)\}$ attached to the particle of sediment
observed today at a point $\r_\tau\in G_\tau$ remain preserved:
                \begin{equation} \label{GBR:ModelPreservation}
                    u(\overline{\r}_\tau) = u({\r}_\tau) \quad ; \quad
                    v(\overline{\r}_\tau) = v({\r}_\tau) \quad ; \quad
                    t(\overline{\r}_\tau) = t({\r}_\tau)
                \end{equation}
As a consequence:
\begin{equation}\label{GBR-is-a-GeoChronModel}
\begin{array}{|c|} \hline
    \mbox{
        \begin{minipage}[t]{130mm}
            \ \\
            Given a present-day GeoChron model of the subsurface characterized
            by GeoChron functions $\{u,v,t\}_\r$, its GBR
            restoration at a given geological-time $\tau$ in the past is also a GeoChron model characterized by the same GeoChron functions.
        \end{minipage}
        }
\\ \\ \hline \end{array}
\end{equation}
\noindent
Therefore, at any restoration time $\tau$, any tool or application developed for
a GeoChron model may be applied as is on the GBR-restored version $\overline{G}_\tau$ of this model.

\ \\
The most important application of geological restoration is to validate the
geometry of the input GeoChron model. At any geological-time $\tau$, this
restored geometry is simpler, which makes validation and editing easier. This
validation process is
robust only if the restoration method is both precise and consistent with the
initial GeoChron model provided as input and we will show how to implement a solution which
addresses these concerns.

\section{Characterizing function $\protect\Bmath{t_\tau(\r_\tau)}$}

\label{BGR:Characterizing-t-tau}

In this section, assume that, at geological-time $\tau$, the effect of
compaction is omitted in $G_\tau$ and that sea floor $\overline{\cal S}_\tau(0)$
coincides with sea level at altitude zero.

\ \\
For any tectonic style, after applying tectonic forces to
jelly block $\overline{G}_\tau$, the images $\{{\cal S}_\tau(d): d\geq 0\}$ of
horizontal surfaces $\{\overline{\cal S}_\tau(d): d\geq 0\}$ remain parallel. For
any $d\geq 0$ and any infinitely small increment $\varepsilon>0$, parallel
surfaces ${\cal S}_\tau(d)$ and ${\cal S}_\tau(d+\varepsilon)$ may be considered
as the top and base of a jelly layer with constant thickness $\varepsilon$. In
other words, for any point $\r_\tau\in{\cal S}_\tau(d+\varepsilon)$, the shortest
path to ${\cal S}_\tau(d)$ measures $\varepsilon$ and is orthogonal to both ${\cal S}_\tau(d)$ and ${\cal S}_\tau(d+\varepsilon)$.

\ \\
As a consequence:
\begin{itemize}

\item Starting from any arbitrary point $\r_\tau\in G_\tau$ there is, recursively defined, a
curvilinear ``normal-line\footnote{ See Equation~\ref{GBR:N}.}''
${\cal N}_\tau(\r_\tau)$ constantly orthogonal to the
family of parallel surfaces $\{{\cal S}_\tau(d):d\geq 0\}$ and linking $\r_\tau$ to the nearest point on $\{{\cal S}_\tau(0)\equiv H_\tau\}$;

\item The value of $t_\tau(\r_\tau)$ is defined as the negative distance along
${\cal N}_\tau(\r_\tau)$ from point $\r_\tau$ to surface $\{{\cal S}_\tau(0)\equiv H_\tau\}$:
        \begin{equation} \label{GBR:normal-line}
        \begin{array}{|c|} \hline \\
                    t_\tau(\r_\tau) \ = \ \Bmath{-}\,
                    \biggl\{\  \mbox{ arc length of normal-line between $\r_\tau$ and ${\cal S}_\tau(0)$}\  \biggr\}
                    \quad \forall\ \r_\tau\in G_\tau
        \\ \\ \hline \end{array}
        \end{equation}
\end{itemize}
\noindent
Moreover, $t_\tau(\r_\tau)$ is also equal to the vertical coordinate $t_\tau(\bar\r_\tau)$ of $\bar\r_\tau$ in $\overline{G}_\tau$.

\ \\
For any derivable function $\varphi(\r)$ and unit vector $\B{u}$, the following
equation holds\footnote{ E.g., see Equation~13.43 on page 316 of
\citet{Mallet2014}.}:
\begin{equation} \label{GBR-JRP-1.0}
    \frac{d\varphi(\r+s\cdot\B{u})}{ds}\bigg|_{s=0}
    \ = \
    \grad\,\varphi(\r)\cdot \B{u}
\end{equation}\noindent
Therefore, if we denote $\B{N}_\tau(\r_\tau)$ the unit vector at location $\r_\tau\in
G_\tau$ which is orthogonal to surface ${\cal S}_\tau(d(\r_\tau))$ passing through
$\r_\tau$ and oriented in the direction of younger terrains, then:
\begin{equation} \label{GBR-JRP-1}
    \frac{d t_\tau(\r_\tau+s\cdot\B{N}_\tau(\r_\tau))}{ds}\bigg|_{s=0}
    \ = \
    \grad\,t_\tau(\r_\tau)\cdot \B{N}_\tau(\r_\tau)
    \ = \
    \grad\,t_\tau(\r_\tau)\cdot \frac{\grad\,t_\tau(\r_\tau)}{||\grad\,t_\tau(\r_\tau)||}
    \ = \ ||\grad\,t_\tau(\r_\tau)||
\end{equation}\noindent
According to Equation~\ref{GBR:normal-line}, $d
t_\tau(\r_\tau+s\cdot\B{N}_\tau(\r_\tau))$ represents the thickness $ds>0$ of
the micro layer between ${\cal S}_\tau(d(\r_\tau))$ and ${\cal
    S}_\tau(d(\r_\tau)-ds)$, from which we can write:
\begin{equation} \label{GBR-JRP-2}
    \{\ dt_\tau(\r_\tau+s\cdot\B{N}_\tau(\r)) \ = \ ds \}
    \quad \Longleftrightarrow \quad
    \biggl\{
    ||\grad\,t_\tau(\r_\tau)||
    \ = \
    \frac{d t_\tau(\r+s\cdot\B{N}_\tau(\r_\tau))}{ds}
    \ = \ 1
    \biggr\}
\end{equation}\noindent
Moreover, on horizon $H_\tau$, we have
\begin{equation} \label{GBR-JRP-2-a}
\forall\ \r_\tau^o\in \{{\cal S}_\tau(0)\equiv H_\tau\} \ : \quad
\left|
\begin{array}{lll}
    1)& t_\tau(\r_\tau^o) = 0
    \\ \\
    2)& \displaystyle
        \B{N}_\tau(\r_\tau^o) = \B{N}(\r_\tau^o)
    \end{array}
\right.
\end{equation}\noindent
where $\B{N}(\r_\tau^o)$, defined by equation \ref{GBR:N}, is given.

\subsection*{The eikonal equation}

\begin{figure}
\centerline{\psfig{figure=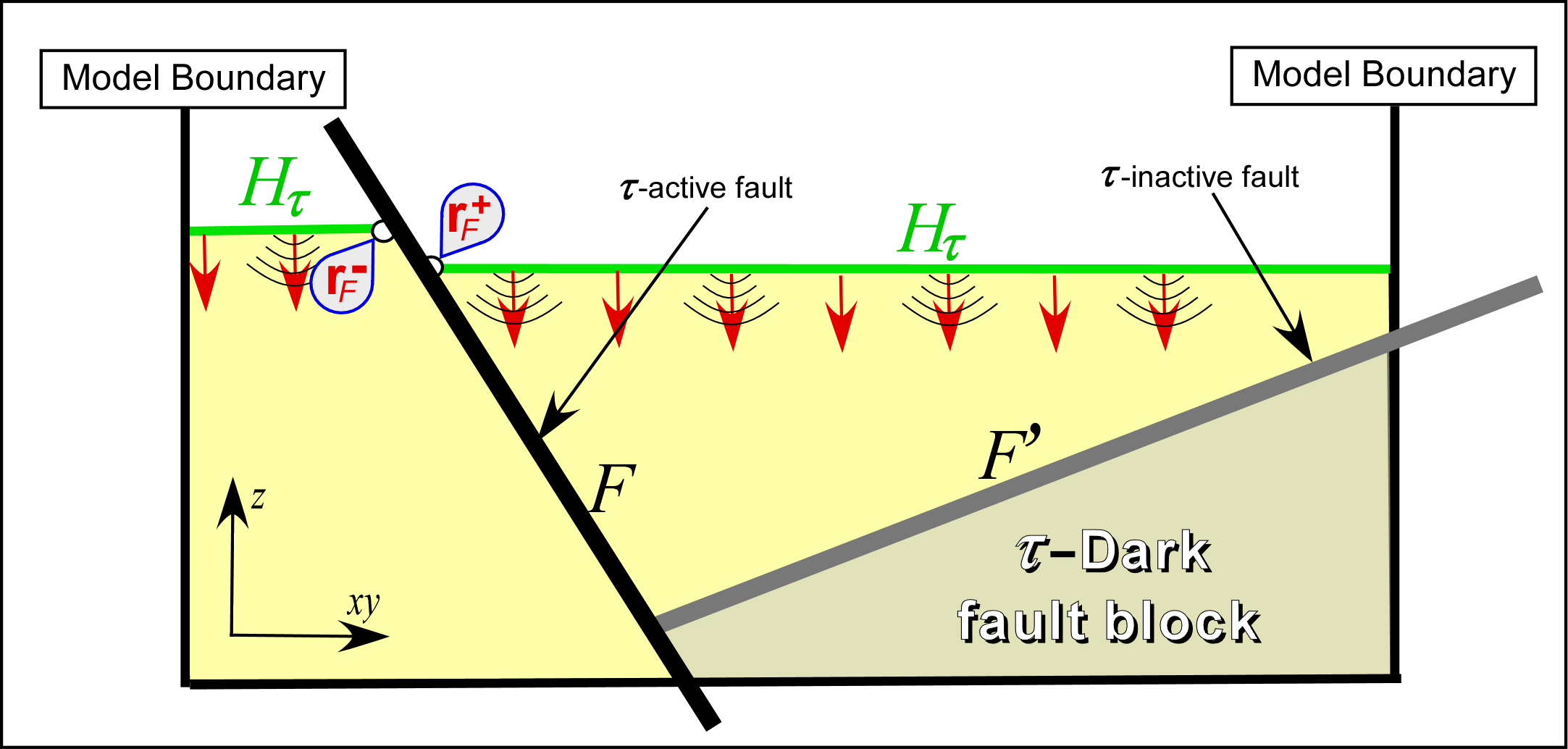,width=116mm}}
\caption{
    Vertical cross section in a structural model depicting a ``$\tau$-dark''
    fault block (darker yellow), which cannot be illuminated by light emitted by horizon $H_\tau$.
        }
\label{GBR-dark-fault-block}
\end{figure}

In a jelly block $G_\tau$ of any geometrical and topological complexity, we may
conclude from the equations above that $t_\tau(\r_\tau)$ must honor the following
fundamental differential equation, called the ``eikonal equation'',
characterizing the parallelism of surfaces $\{{\cal S}_\tau(d):d\geq 0\}$,
subject to specific boundary conditions:
        \begin{equation} \label{GBR:eikonal-equation}
        \begin{array}{|c|} \hline \\
              \begin{array}{cccc}
                    1)& ||\grad\,t_\tau(\r_\tau)|| = 1 \qquad \forall\ \r_\tau\in G_\tau
                    \\ \\
                    2)& \mbox{subject to : \ }
                        \left\{
                        \begin{array}{cc}
                            a)&   t_\tau(\r_\tau^o) = 0
                            \\ \\
                            b)&  \displaystyle
                            \grad\,t_\tau(\r_\tau^o) = \B{N}(\r_\tau^o)
                        \end{array}
                        \right\} \quad \forall\ \r_\tau^o\in \{{\cal S}_\tau(0)\equiv H_\tau\}
              \end{array}\noindent
        \\ \\ \hline \end{array}
        \end{equation}\noindent
Physicists would use the well-known eikonal Equation~\ref{GBR:eikonal-equation} to
describe the time of first arrival at point $\r_\tau\in G_\tau$ of a light wave-front
emitted by $\{{\cal S}_\tau(0)\equiv H_\tau\}$ and propagating at constant, unit
speed. To go further with this analogy, faults would be considered as opaque barriers which induce
discontinuities in functions $\{u_\tau,v_\tau,t_\tau\}_{\r_\tau}$. As
Figure~\ref{GBR-dark-fault-block} shows, fault blocks which are not illuminated by
$H_\tau$ are called ``$\tau$-dark fault blocks''. More precisely, a point
$\r_\tau\in G_\tau$ belongs to a $\tau$-dark fault block if and only if, within the
studied domain, no continuous path (i.e. uncut by faults) exists between $\r_\tau$ and $\{{\cal S}_\tau(0)\equiv H_\tau\}$.

\noindent

\section{Characterizing functions $\protect\Bmath{\{u_\tau, v_\tau\}_{\r_\tau}}$}

\label{BGR:Characterizing-uv-tau}

Solving eikonal Equation~\ref{GBR:eikonal-equation} provides us with the values
of $t_\tau(\r_\tau)$ over space $G_\tau$. Assuming that $t_\tau(\r_\tau)$ is now
known, this section shows how differential
equations characterizing functions $\{u_\tau, v_\tau\}_{\r_\tau}$ can be derived from
the jelly paradigm and fundamental GBR principle~\ref{GBR-fundamental-principle}.

\subsection*{First type boundary conditions for $\protect\Bmath{\{u_\tau,v_\tau\}_{\r_\tau}}$ on $\protect\Bmath{H_\tau}$}

By definition, for any point $\r_\tau^o\in H_\tau$:
\begin{itemize}

\item In the $G_\tau$ space, $\{u(\r_\tau^o),v(\r_\tau^o)\}$ are the given GeoChron paleo-geographic coordinates of the particle of sediment observed today at location $\r_\tau^o$;

\item In the $\overline{G}_\tau$ restored space,
    $\{u_\tau(\bar\r_\tau^o),v_\tau(\bar\r_\tau^o)\}$ are unknown geographic coordinates
    of the particle of sediment which would have been observed at location $\bar\r_\tau^o$ on sea floor $\overline{\cal S}_\tau(0)$.

\end{itemize}
\noindent
Obviously, taking Equations~\ref{GBR-XXX-1} into account and for any point
$\r_\tau^o\in H_\tau$, coordinates
$\{u(\r_\tau^o),v(\r_\tau^o)\}$ and $\{u_\tau(\r_\tau^o),v_\tau(\r_\tau^o)\}$
should be identical. As a consequence, the following first type boundary
conditions, where $u(\r_\tau^o)$ and $v(\r_\tau^o)$ are known, must be honored:
    \begin{equation} \label{GBR:Indet}
    \begin{array}{|c|} \hline \\
        \forall\ \r_\tau^o\in \{{\cal S}_\tau(0)\equiv H_\tau\} \ : \quad
        \left|
        \begin{array}{cccl}
        1)& u_\tau(\r_\tau^o) &=& u(\r_\tau^o)
        \\ \\
        2)& v_\tau(\r_\tau^o)&=& v(\r_\tau^o)
        \end{array}
        \right.
    \\ \\ \hline \end{array}
    \end{equation}\noindent
\citet{Lovely2018} do not set the constraints specified above, which
implies that functions $\{u_\tau,v_\tau\}_{\r_\tau}$ are not synchronized with known
paleo-geographic functions $\{u,v\}_\r$. As a consequence, the $uvt$-transform
and $u_\tau v_\tau t_\tau$-transform of $H_\tau$ are not constrained to be
identical, implying that erroneous deformations may appear on
$\overline{H}_\tau$ obtained as $u_\tau v_\tau t_\tau$-transform of $H_\tau$. As
an example, consider the (generally curvilinear) patch $H_\tau(u_o,v_o,\Delta)$
defined on $H_\tau$ as follows:
        \begin{equation} \label{eqn:Restoration-Rescal-4}
        \r\in H_\tau(u_o,v_o,\Delta)
        \quad \Longleftrightarrow \quad
        \left\{
        \begin{array}{ccccc}
           t(\r)=\tau
           \\
           u_o\leq u(\r) \leq u_o+\Delta
           \\
           v_o\leq v(\r) \leq v_o+\Delta
        \end{array}
        \right.
        \end{equation}\noindent
and consider also $\overline{H}_\tau^\oplus(u_o,v_o,\Delta)$ and
$\overline{H}_\tau^\ominus(u_o,v_o,\Delta)$ as the restored images of
$H_\tau(u_o,v_o,\Delta)$ on $\overline{H}_\tau$ with and without constraints~\ref{GBR:Indet},
respectively. If
constraints~\ref{GBR:Indet} are omitted, then
$\overline{H}_\tau^\oplus(u_o,v_o,\Delta)$ and $\overline{H}_\tau^\ominus(u_o,v_o,\Delta)$
may have different areas and/or shapes, implying that the restoration
process may induce deformations which are incoherent with
those described by the $\{u,v,t\}_\r$ GeoChron functions provided as input.

\subsection*{Second type boundary conditions for $\protect\Bmath{\{u_\tau,v_\tau\}_{\r_\tau}}$ on $\protect\Bmath{H_\tau}$}

In the restored space $\overline{G}_\tau$, terrains older than $\tau$ are
generally still folded and, similarly to terrains in the $G$-space, their
deformation may be characterized by the ``partial'' strain tensor at geological-time $\tau$ denoted $\Bmath{\cal E}(\r|\tau)$.
For coherency's sake, on $H_\tau$ the ``total'' strain tensor $\Bmath{\cal
E}(\r)$ characterized by Equation~2.20 on page~63 of \citet{Mallet2014} and the ``partial'' strain tensor $\Bmath{\cal E}(\r|\tau)$ should be equal:
    \begin{equation} \label{GBR:BCUV-2T.X1}
       \Bmath{\cal E}(\r_\tau^o) \ = \ \Bmath{\cal E}(\r_\tau^o|\tau)
       \qquad\forall\ \r_\tau^o\in H_\tau
    \end{equation}\noindent
On horizon $H_\tau$, as a consequence of boundary conditions
\ref{GBR:eikonal-equation}.2b set on $t_\tau(\r_\tau)$, we have:
    \begin{equation} \label{GBR:BCUV-2T.0}
       \B{N}(\r_\tau^o) \ = \ \frac{\grad\,t(\r_\tau^o)}{||\grad\,t(\r_\tau^o)||} \ = \  \frac{\grad\,t_\tau(\r_\tau^o)}{||\grad\,t_\tau(\r_\tau^o)||} \ = \ \B{N}_\tau(\r_\tau^o)
       \qquad\forall\ \r_\tau^o\in H_\tau
    \end{equation}\noindent
Therefore, barring the effects of compaction,
according to GBR principle~\ref{GBR-fundamental-principle} and
Equation~\ref{GBR:BCUV-2T.X1}, for all indexes $(\alpha,\beta)\in\{x,y,z\}^2$, Equation~\ref{eqn:Strain-6.4} implies:
    \begin{equation} \label{GBR:BCUV-2T.X2}
    \forall\ \r_\tau^o\in H_\tau \ : \quad
    \left|
    \begin{array}{llllllllll}
    &&
               \{
                    \partial_\alpha u\cdot\partial_\beta u + \partial_\alpha v\cdot\partial_\beta v + {N^\alpha\cdot N^\beta }
               \}_{\r_\tau^o}
    \\ \\
    &=&
               \{
                    \partial_\alpha u_\tau\cdot\partial_\beta u_\tau + \partial_\alpha v_\tau\cdot\partial_\beta v_\tau + {N^\alpha\cdot N^\beta }
               \}_{\r_\tau^o}
    \end{array}
    \right.
    \end{equation}\noindent
    \begin{equation} \label{GBR:BCUV-2T.4}
    \Longleftrightarrow\quad
    \forall\ \r_\tau^o\in H_\tau \ : \quad
    \left|
    \begin{array}{cllllll}
    &&
               \{
                    \partial_\alpha u\cdot\partial_\beta u + \partial_\alpha v\cdot\partial_\beta v
               \}_{\r_\tau^o}
    \\ \\
    &=&
               \{
                    \partial_\alpha u_\tau\cdot\partial_\beta u_\tau + \partial_\alpha v_\tau\cdot\partial_\beta v_\tau
               \}_{\r_\tau^o}
    \end{array}
    \right.
    \end{equation}\noindent
A straightforward solution to these equations consists in constraining
restoration functions $\{u_\tau,v_\tau\}_{\r_\tau}$ as follows, where
$\{u,v\}_\r$ are known:
    \begin{equation} \label{GBR:Indet-1}
    \begin{array}{|c|} \hline \\
        \forall\ \r_\tau^o\in \{{\cal S}_\tau(0)\equiv H_\tau\} \ : \quad
        \left|
        \begin{array}{cccl}
        1)& \grad\,u_\tau(\r_\tau^o) &=& \grad\,u(\r_\tau^o)
        \\ \\
        2)& \grad\,v_\tau(\r_\tau^o)&=& \grad\,v(\r_\tau^o)
        \end{array}
        \right.
    \\ \\ \hline \end{array}
    \end{equation}\noindent
This second type of boundary conditions are not implemented in
\citet{Lovely2018}'s method, which may jeopardize the consistency of
restored models with respect to the initial GeoChron model.

\subsection*{Comment}

In order to maintain consistency of restoration functions
$\{u_\tau,v_\tau,t_\tau\}_{\r_\tau}$ with input GeoChron functions
$\{u,v,t\}_{\r_\tau}$, boundary conditions~\ref{GBR:Indet} and~\ref{GBR:Indet-1}
must be honored as strictly as possible. If they do not conflict with these
boundary conditions, other, not so strict constraints may be added and honored
in a least squares sense, as for example the following ``$\tau$-twin-pins''
constraint.

\ \\
By definition, we suggest denoting ``$\tau$-twin-pins'' a pair of points
$(\r_\tau^1,\r_\tau^2)$ in $G_\tau$ such that the structural geologist has
reason to believe that their restored images
$(\overline{\r}_\tau^1,\overline{\r}_\tau^2)$ at restoration time $\tau$ must be
located on a single vertical line in $\overline{G}_\tau$. In order to take that
constraint into account, restoration functions
$\{u_\tau,v_\tau,t_\tau\}_{\r_\tau}$ may be computed such that, in a least
squares sense:
\begin{equation} \label{GBR:Pin-Point}
    \left|
    \begin{array}{ccccccccc}
        u_\tau(\r_\tau^1) &\simeq& u_\tau(\r_\tau^2)
        \\ \\
        v_\tau(\r_\tau^1) &\simeq& v_\tau(\r_\tau^2)
    \end{array}
    \right.
\end{equation}\noindent
By setting this type of constraint repeatedly on pairs of points located, for
instance, on a line in $G_\tau$, it is possible to make the restored version of
this line vertical in $\overline{G}_\tau$.

\subsection*{Characterizing functions $\protect\Bmath{\{u_\tau,v_\tau\}_{\r_\tau}}$ in $G_\tau$}

According to GeoChron theory\footnote{ See equation \ref{eqn:Strain-6.4}.}, everywhere in studied domain $G$, terrain
deformation is characterized by the gradients of geological-time function
$t(\r)$ and paleo-geographic functions $\{u,v\}_\r$, which may be considered as
deformation ``records'' taken into account by boundary
conditions~\ref{GBR:Indet} and~\ref{GBR:Indet-1}. As explained below, to
propagate these boundary conditions over the entire $G_\tau$-space, the
gradients of $\{u_\tau,v_\tau\}_{\r_\tau}$ have to honor specific differential
equations.

\ \\
Referring to fundamental GBR principle~\ref{GBR-fundamental-principle}, to
characterize the restoration functions $\{u_\tau,v_\tau,t_\tau\}_\r$, we should
simply substitute these functions to $\{u,v,t\}_\r$ in
Equation~\ref{eqn:geochron:43E3} or Equation~\ref{eqn:geochron:49e}. However:
\begin{enumerate}

\item In $G_\tau$ and in accordance with GeoChron theory:
    \begin{enumerate}

        \item ``soft'' constraints~\ref{eqn:geochron:43E3}-1\&2
        or~\ref{eqn:geochron:49e}-1\&2 may only be honored in a least squares
        sense;

        \item due to local deformations of horizons and layers induced by
        tectonic forces, left hand sides of
        constraints~\ref{eqn:geochron:43E3}-1\&2 or~\ref{eqn:geochron:49e}-1\&2
        may slightly differ from the specified value ``1''.

    \end{enumerate}\noindent

\item On $H_\tau$, ``hard'' constraints~\ref{GBR:Indet} and~\ref{GBR:Indet-1}
strictly specify the values of restoration functions $\{u_\tau,v_\tau\}$ and
their 3D gradients which, according to item 1.b above, may conflict with
constraints~\ref{eqn:geochron:43E3}-1\&2 or~\ref{eqn:geochron:49e}-1\&2.

\end{enumerate}\noindent
Therefore, to resolve such a conflict, we suggest removing
constraints~\ref{eqn:geochron:43E3}-1\&2 or~\ref{eqn:geochron:49e}-1\&2
and constraining $\{u_\tau,v_\tau\}_{\r_\tau}$ in a least squares sense as follows:
\begin{itemize}

\item In a minimal deformation tectonic style context:
            \begin{equation} \label{GBR:X11}
            \forall\ \r_\tau\in G_\tau \ : \quad
            \begin{array}{|c|} \hline \\
                  \begin{array}{cc}
                  1)&
                  \{\grad\,u_\tau \cdot \grad\,v_\tau\}_{\r_\tau} \ \simeq \ 0
                  \\ 
                  2)&
                  \{\grad\,t_\tau \cdot \grad\,u_\tau\}_{\r_\tau} \ \simeq \ 0
                  \\ 
                  3)&
                  \{\grad\,t_\tau \cdot \grad\,v_\tau\}_{\r_\tau} \ \simeq \ 0
                  \end{array}
            \\ \\ \hline \end{array}
            \end{equation}

\item In a flexural slip tectonic style context, denoting
$\grad_{\msc{s}}\varphi(\r_\tau)$ the orthogonal projection of $\grad\,\varphi(\r_\tau)$ onto
the plane tangent to the level-set ${\cal S}_\tau(-t_\tau(\r_\tau))$ of function
$t_\tau(\r_\tau)$ at location $\r_\tau$:
            \begin{equation} \label{GBR:S1}
            \forall\ \r_\tau\in G_\tau\ : \quad
            \begin{array}{|c|} \hline \\
                        \{\grad_{\msc{s}}\,u_\tau \cdot \grad_{\msc{s}}\,v_\tau\}_{\r_\tau} \ \simeq \ 0
            \\ \\ \hline \end{array} 
            \end{equation}\noindent

\end{itemize}
It must be noted, however, that boundary conditions~\ref{GBR:Indet}
and~\ref{GBR:Indet-1} fully specify $\{u_\tau,v_\tau\}_{\r_\tau}$ only on
horizon $H_\tau$. To propagate these conditions downward throughout the
whole domain $G_\tau$ whilst honoring constraints \ref{GBR:X11} or \ref{GBR:S1}, we propose to specify that the gradients of these
functions must vary as smoothly as possible in $G_\tau$. In practice, this may
be achieved in a least squares sense thanks to the following, additional constraint:
                \begin{equation} \label{GBR-UtauVtau-smoothness}
                 \sum_{\alpha\in\{x,y,z\}}
                                \int_{G_\tau}
                                \biggl[
                                    || \partial_\alpha \grad\,u_\tau(\r_\tau) ||^2  +
                                    || \partial_\alpha \grad\,v_\tau(\r_\tau)||^2
                                \biggr] \cdot d\r_\tau \qquad \mbox{minimum}
                \end{equation}\noindent

\section{Accounting for faults}

\label{GBR-taking-faults-into-account}

As shown on Figure~\ref{GBR-Tetrahedral_Mesh-1}, 3D geological domain $G$ may be
cut by faults and GeoChron functions $\{u,v,t\}_\r$ are discontinuous
across these faults. Similarly, faults induce discontinuities in GeoChron
functions $\{u_\tau,v_\tau,t_\tau\}_{\r_\tau}$ defined on $G_\tau$. However, based on
geological arguments presented below, values and gradients of functions
$\{u_\tau,v_\tau,t_\tau\}_{\r_\tau}$ on either side of a fault should generally honor geometric constraints which are
specific to particular types of faults.

\subsection*{$\Bmath{\tau}$-active \& $\Bmath{\tau}$-inactive faults}

\begin{figure}
\centerline{\psfig{figure=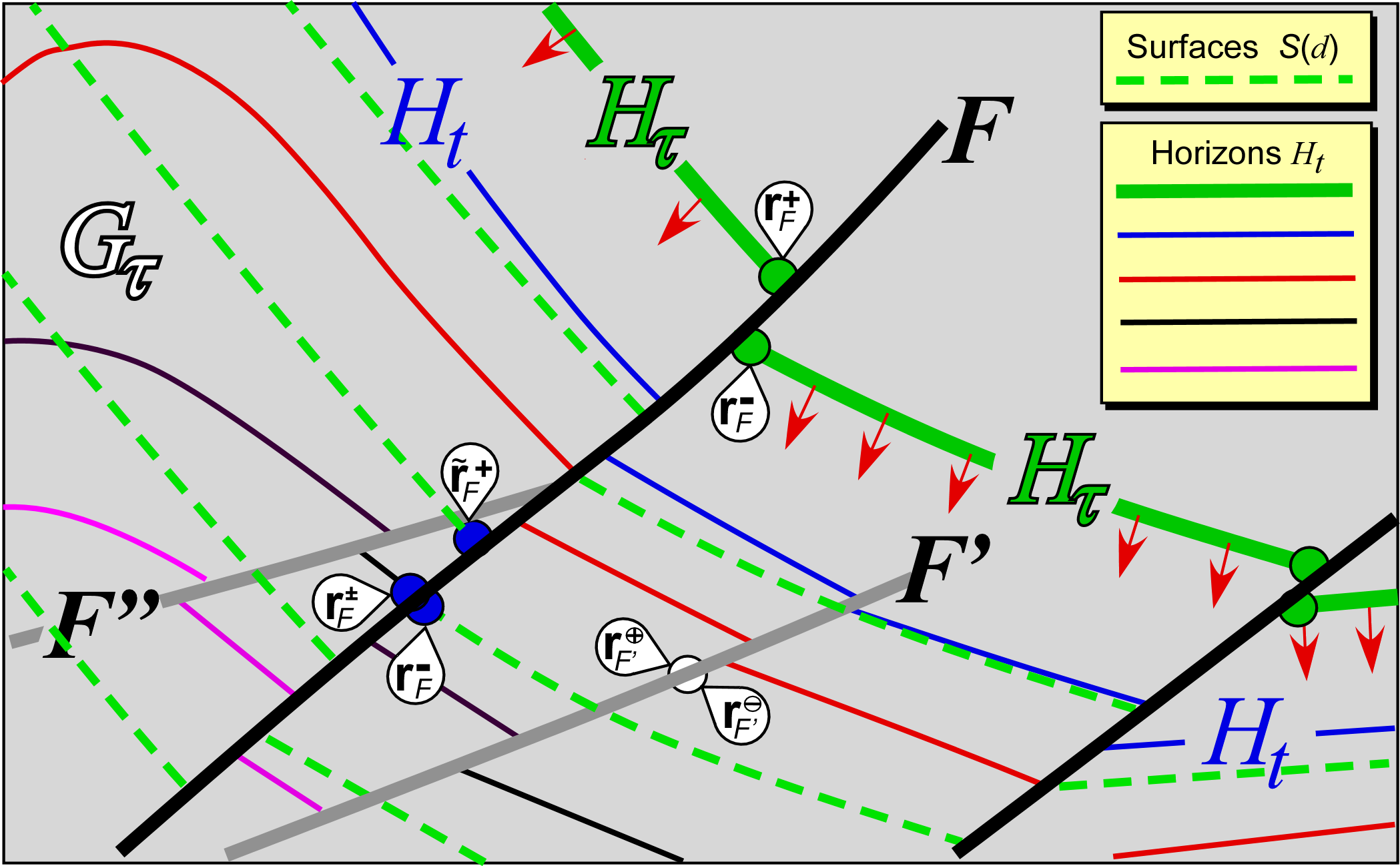,width=116mm}}
\caption{
    Vertical cross section where multicolored bold lines represent horizons
    $\{H_t: t\leq\tau\}$ and green dashed lines depict surfaces $\{{\cal S}_\tau(d):d\geq 0\}$ parallel to $\{H_\tau\equiv{\cal S}_\tau(0)\}$. For each pair of
twin-points $(\r_{\msc{f}}^+,\r_{\msc{f}}^-)$ on $H_\tau$, we have both
$\{t_\tau(\r_{\msc{f}}^+)=t_\tau(\r_{\msc{f}}^-)=z_\tau^o\}$ and
$\{t(\r_{\msc{f}}^+)=t(\r_{\msc{f}}^-)\}$. Level sets $\{{\cal S}_\tau(d):d\geq 0\}$ of
function $t_\tau(\r_\tau)$ are continuous across $\tau$-inactive faults (in gray).
        }
\label{GBR-DyingFaults}
\end{figure}

With respect to a given restoration time $\tau$, we classify faults according to the
two following categories:
\begin{itemize}
            \item A fault which intersects horizon $H_\tau$ is a
``$\tau$-active'' fault (e.g $F$ in Figure~\ref{GBR-DyingFaults});

            \item A fault which belongs to $G_\tau$ and does not intersect
horizon $H_\tau$ is a ``$\tau$-inactive'' fault (e.g.
$F^{\prime}$ and $F^{\prime\prime}$ in Figure~\ref{GBR-DyingFaults}).

\end{itemize}
\noindent
``$\tau$-active'' or ``$\tau$-inactive'' status is defined relatively to
restoration time $\tau$: At an older restoration time
$\tau^\prime<\tau$, a $\tau$-inactive fault which intersects $H_{\tau^\prime}$
may become a $\tau^\prime$-active fault.
\noindent

\ \\
When horizon $H_\tau$ is restored, the jelly block $G_\tau$
underneath $H_\tau$ must behave as if it were only impacted by $\tau$-active
faults. All other geologic objects such as horizons and $\tau$-inactive faults
embedded in the jelly block are passively deformed by this restoration
process. In other words, restoration functions
$\{u_\tau,v_\tau,t_\tau\}_{\r_\tau}$ must be continuous across $\tau$-inactive faults.

\ \\
However, geological domain $G_\tau$ is topologically
discontinuous across faults of any type.
In order to ensure functions $\{u_\tau,v_\tau,t_\tau\}_{\r_\tau}$ are ${\cal C}^1$-continuous across
$\tau$-inactive faults, the following constraints may be set on pairs of
``$\tau$-mate-points'' $(\r_{\msc{f}}^\oplus,\r_{\msc{f}}^\ominus)_\tau$
defined as collocated points lying on $F^+$
and $F^-$, respectively:
            \begin{equation} \label{GBR-Secondary-faults}
            \begin{array}{|c|} \hline \\
                  \left|
                  \begin{array}{cccccccccccccccccc}
                    1)& u_\tau(\r_{\msc{f}}^\oplus) &=& u_\tau(\r_{\msc{f}}^\ominus)
                    \\
                    2)& v_\tau(\r_{\msc{f}}^\oplus) &=& v_\tau(\r_{\msc{f}}^\ominus)
                    \\
                    3)& t_\tau(\r_{\msc{f}}^\oplus) &=& t_\tau(\r_{\msc{f}}^\ominus)
                  \end{array}
                  \right.
                  \quad \Bmath{\&} \quad
                  \left|
                  \begin{array}{cccccccccccccccccc}
                    4)& \grad\,u_\tau(\r_{\msc{f}}^\oplus) &=& \grad\,u_\tau(\r_{\msc{f}}^\ominus)
                    \\
                    5)& \grad\,v_\tau(\r_{\msc{f}}^\oplus) &=& \grad\,v_\tau(\r_{\msc{f}}^\ominus)
                    \\
                    6)& \grad\,t_\tau(\r_{\msc{f}}^\oplus) &=& \grad\,t_\tau(\r_{\msc{f}}^\ominus)
                  \end{array}
                  \right.
                \\ \\
                    \forall\ (\r_{\msc{f}}^\oplus,\r_{\msc{f}}^\ominus)_\tau \in \ \mbox{$\tau$-inactive fault $F$}
            \\ \\ \hline \end{array}
            \end{equation}
            \noindent

\subsection*{Boundary conditions for $\protect\Bmath{\{u_\tau,v_\tau\}_{\r_\tau}}$ on $\protect\Bmath{\tau}$-active faults}

When horizon $H_\tau$ is restored, terrains located on either side of
$\tau$-active faults must slide along lines called $\tau$-fault-striae, tangent
to these faults. For geological consistency, $\tau$-fault-striae must be
identical to fault-striae (see Figure~\ref{GBR-fault-striae}) associated with the $\{u,v\}_\tau$
paleo-geographic coordinates of the GeoChron model provided as input to the
proposed restoration method:
        \begin{equation} \label{GBR-fault-striae-identity}
        \begin{array}{|c|} \hline \\
            \Bmath{\tau}\mbox{-fault-striae}\ \equiv \ \mbox{fault-striae}
            \qquad \forall\ \tau
        \\ \\ \hline \end{array}
        \end{equation}\noindent
        \noindent
As \citet{Lovely2018} use the standard
SKUA\textsuperscript{\tiny\textregistered} algorithm to reestablish continuity
of functions $\{u_\tau,v_\tau\}_{\r_\tau}$ through faults, $\tau$-twin-points are
recomputed from the geometry and
topology of level-sets $\{{\cal S}(d):d\geq 0\}$ of $t_\tau(\r_\tau)$ in $G_\tau$. As
a consequence, these $\tau$-twin-points may not be located on the same
fault-striae as those induced by known $\{u,v,t\}_\r$ functions of the input
GeoChron model (e.g., see Figure~\ref{GBR-fault-striae}), which may break the
consistency between input and restored model close to faults.

\ \\
Similarly to GeoChron twin-points $(\r_{\msc{f}}^+,\r_{\msc{f}}^-)$\footnote{ See
definition~\ref{GBR:TP-1}.}, $\tau$-twin-points
$(\r_{\msc{f}}^+,\r_{\msc{f}}^-)_\tau$ are characterized by the following
equations:
            \begin{equation} \label{GBR:TP-1.tau}
                        \{\ (\r_{\msc{f}}^+,\r_{\msc{f}}^-)_\tau \mbox{ is a pair of $\tau$-twin-points } \}
                        \quad \Longleftrightarrow \quad
                        \left\{
                        \begin{array}{ccccccc}
                            1)& \mbox{$F$ is a $\tau$-active fault} \\
                            2)& \r_{\msc{f}}^+ \in F^+ \ \mbox{ \B{\&} } \ \r_{\msc{f}}^- \in F^- \\ \\
                            3)&  u_\tau(\r_{\msc{f}}^-) = u_\tau(\r_{\msc{f}}^+)  \\
                            4)&  v_\tau(\r_{\msc{f}}^-) = v_\tau(\r_{\msc{f}}^+)  \\
                            5)&  t_\tau(\r_{\msc{f}}^-) = t_\tau(\r_{\msc{f}}^+)
                        \end{array}
                        \right.
            \end{equation}\noindent


\ \\
In order to restore terrains along $\tau$-active faults without creating voids
or overlaps,
constraints~\ref{GBR:TP-1.tau}-3 to \ref{GBR:TP-1.tau}-5 must be honored
all along $\tau$-active faults.
Function $t_\tau(\r_\tau)$ is independent from $\{u_\tau,v_\tau\}_{\r_\tau}$ and
is assumed to be already known. As a consequence, $\{u_\tau,v_\tau\}_{\r_\tau}$ simply have to honor the
following constraints:
\begin{equation} \label{GBR-Sigma-6}
\begin{array}{|c|} \hline \\
\left.
\begin{array}{ccccccccccccc}
   1)& u_\tau(\r_{\msc{f}}^+) &=& u_\tau(\r_{\msc{f}}^-)
   \\ \\
   2)& v_\tau(\r_{\msc{f}}^+) &=& v_\tau(\r_{\msc{f}}^-)
\end{array}
\right\}
\quad \forall\ (\r_{\msc{f}}^+,\r_{\msc{f}}^-)_\tau \in \mbox{ $\tau$-active fault}
\\ \\ \hline \end{array}
\end{equation}\noindent

\subsection*{Why distinguish $\Bmath{\tau}$-active from $\Bmath{\tau}$-inactive faults?}

\begin{figure}
\centerline{\psfig{figure=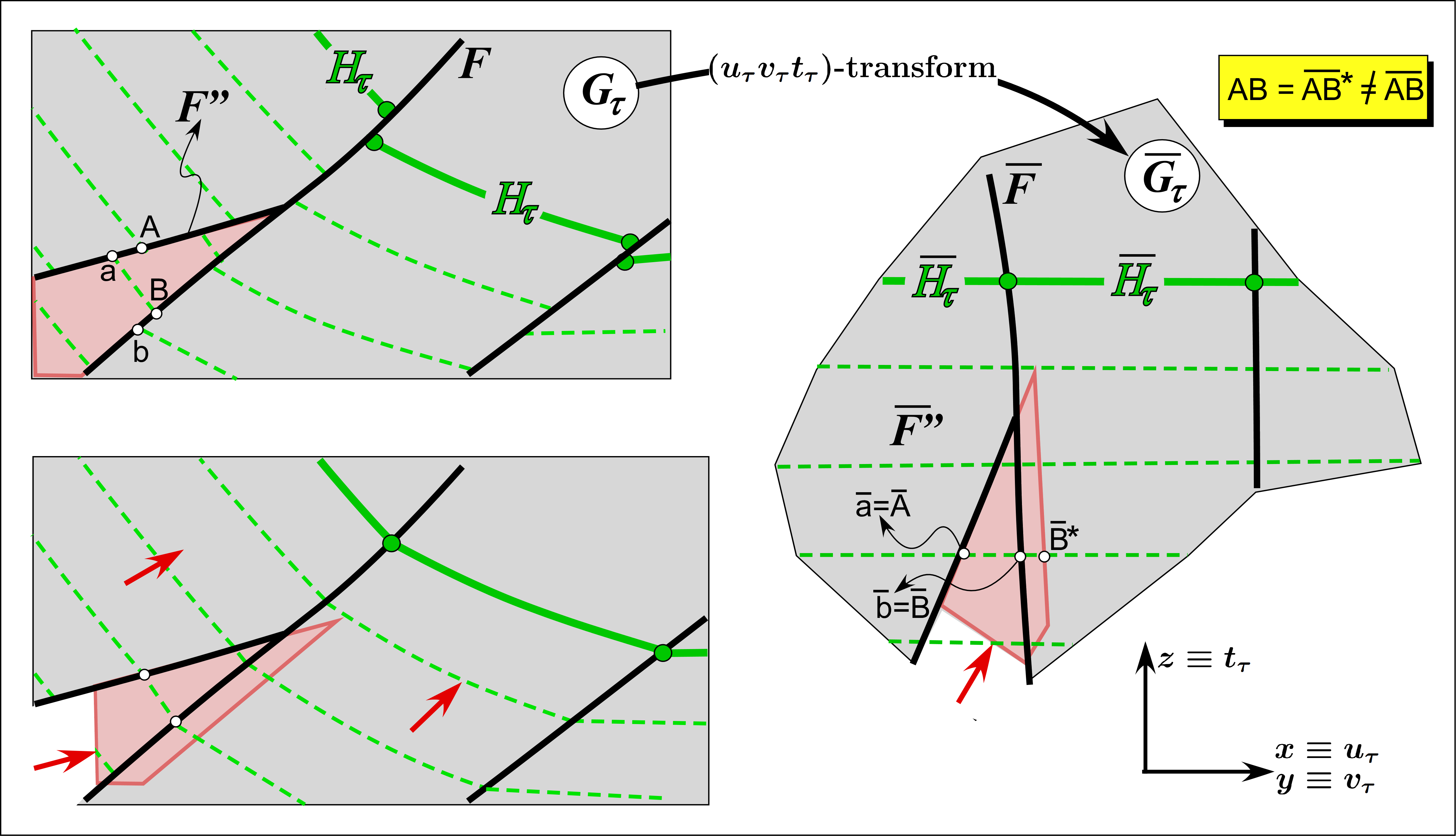,width=120mm}}
\caption{
Vertical cross section in which fault $F^{\prime\prime}$ is
erroneously considered as a $\tau$-active fault. In the restoration process
induced by a $u_\tau v_\tau t_\tau$-transform, the distance between points
$\B{A}$ and $\B{B}$ cannot be correctly preserved.
        }
\label{GBR-DyingFaults-1}
\end{figure}

At restoration geological-time $\tau$, any fault that isolates a fault block in
$G_\tau$ such that pseudo light emitted by $H_\tau$ cannot reach it, must be
considered as $\tau$-inactive in order for this fault block to be restored.

\ \\
Next, considering faults that do not intersect $H_\tau$ as active may result in
erroneous distortion of restored terrains. The top left corner of
Figure~\ref{GBR-DyingFaults-1} shows the same cross section as
Figure~\ref{GBR-DyingFaults} but the restoration for horizon
$H_\tau$ is computed while considering fault $F^{\prime\prime}$, which does not
intersect $H_\tau$, as a $\tau$-active fault.

\ \\
\noindent
If fault $F^{\prime\prime}$ is considered as a $\tau$-active fault, pairs of
points $(\B{a},\B{A})$ and $(\B{b},\B{B})$ shown in the top left part of
Figure~\ref{GBR-DyingFaults-1} are considered as $\tau$-twin-points.
After restoration, $\tau$-twin-points are collocated, implying that
$(\overline{\B{a}}=\overline{\B{A}})$ and $(\overline{\B{b}}=\overline{\B{B}})$
(right hand side of Figure~\ref{GBR-DyingFaults-1}).

\ \\
As distance $(\B{AB}=\overline{\B{AB}^\star})$ differs from
$(\overline{\B{AB}}=\overline{\B{ab}})$, in the neighborhood
of faults $F$ and $F^{\prime\prime}$, a restoration
performed via $u_\tau v_\tau t_\tau$-transform would generate incorrect
variations in lengths and volumes.

\ \\
To avoid these inconsistencies,
distinguishing $\tau$-active and $\tau$-inactive faults is a key component of
the GBR method and an improvement over the first implementation of a GeoChron
restoration method by \citet{Lovely2018}.

\subsection*{Manually activating faults}

According to geological context, structural geologists may prefer some
$\tau$-inactive faults to be considered as $\tau$-active, for example when a
thrust fault is known to be active at a particular time even though it did not
break through to the sea floor. Technically, this is possible for nearly any
fault, which is then constrained with Equations~\ref{GBR-Sigma-6} as any other
active fault.

\ \\
However, any fault bordering a $\tau$-dark fault block (e.g.
Figure~\ref{GBR-dark-fault-block}) must be handled as $\tau$-inactive, otherwise
functions $\{u_\tau,v_\tau,t_\tau\}_{\r_\tau}$ would be undetermined inside this
$\tau$-dark fault block which, as a consequence, could not be restored.

\section{Taking compaction into account}

\label{GBR:::compaction}

Compaction is defined as pore space reduction in sediments due to increased load
during deposition. As this process changes the geometry of geological
layers as their depths increase, restoration workflows frequently handle
compaction as an option.

\ \\
As we have done so far, assume that an initial version of restoration functions
$\{u_\tau,v_\tau,t_\tau\}_{\r_\tau}$ has been obtained without taking compaction
into account. In other words, the compacted thicknesses of layers in studied domain
$G_\tau$, as observed today, have been approximately preserved in restored domain
$\overline{G}_\tau$. In effect, restoration uplifts and unloads terrains, which
should induce decompaction resulting in increased layer thicknesses in the restored domain.

\ \\
In this section we show how $\{u_\tau,v_\tau,t_\tau\}_{\r_\tau}$ can be replaced
by new restoration functions in such a way that
the new $u_\tau v_\tau t_\tau$-transform
$\overline{G}_\tau$ of $G_\tau$ so obtained restores the terrains and induces
thickness variations as a consequence of decompaction, which should be the
exact inverse of the compaction that occurred between geological-time $\tau$ and
the present geological-time.

\subsection*{Athy's law}

As the concept of decompaction may be easier to grasp in the restored space, we
refer to Figure~\ref{GBRestoration-1}-D which shows the subsurface restored at geological-time
$\tau$. Let $\overline{V}(\overline{\r}_\tau)$ be an infinitely small
volume of sediment centered on point $\overline{\r}_\tau\in\overline{G}_\tau$
underneath the sea floor $\{\overline{\cal S}_\tau(0)\equiv\overline{H}_\tau\}$.

\ \\
Laboratory experiments on rock samples show that, during burial when sediments
contained in $\overline{V}(\overline{\r}_\tau)$
compact under their own weight, their porosity $\overline\Psi(\overline{\r}_\tau)$
exponentially decreases according to Athy's law \citep{Athy1930}:
    \begin{equation} \label{GBR::Compaction-0}
        \overline\Psi(\overline{\r}_\tau) \ \simeq \ \overline\Psi_o(\overline{\r}_\tau) \cdot exp\biggl\{ -\overline\kappa(\overline{\r}_\tau)\cdot\delta(\overline{\r}_\tau) \biggr\}
        \qquad \forall\ \overline{\r}_\tau\in \overline{G}_\tau
    \end{equation}\noindent
In this equation, $\delta(\overline{\r}_\tau)$ is
the absolute distance, or depth, from point
$\overline{\r}_\tau\in\overline{G}_\tau$ to sea floor $\overline{\cal
    S}_\tau(0)$ measured at geological-time $\tau$ whilst
    $\overline\Psi_o(\overline{\r}_\tau)<1$ and
    $\overline\kappa(\overline{\r}_\tau)>0$ are known non-negative coefficients
    which only depend on rock type at location $\overline{\r}_\tau$.
As an example, assuming that $\delta(\overline{\r}_\tau)$ is expressed in
meters, the following average coefficients for sedimentary terrains observed
in southern Morocco have been reported \citep{Labbassi1999}:
\begin{center}
   \begin{tabular}{|l|c|c|}
     \hline
     $Rock\ type$ & $\overline\Psi_o$ & $\overline\kappa$ \\
     \hline
         Siltstone & 0.62 & 0.57$\times 10^{-3}$\\
         Clay & 0.71& 0.77$\times 10^{-3}$\\
         Sandstone & 0.35& 0.60$\times 10^{-3}$\\
         Carbonates & 0.46& 0.23$\times 10^{-3}$\\
         Dolomites & 0.21&   0.61$\times 10^{-3}$\\
     \hline
   \end{tabular}
\end{center}
\ \\
Keeping in mind that, in the restored $\overline{G}_\tau$-space,
$\{-{t}_\tau(\overline{\r}_\tau)\}$ measures the vertical distance from
$\overline{\r}_\tau$ to sea floor $\overline{S}_\tau(0)$, in
Equation~\ref{GBR::Compaction-0}, depth function
$\delta(\overline{\r}_\tau)$ can be expressed as follows:
\begin{equation} \label{GBR:decompaction-D-1}
      \delta(\overline{\r}_\tau) \ = \ - {t}_\tau(\overline{\r}_\tau)
      \qquad \forall\ \overline{\r}_\tau \in \overline{G}_\tau
\end{equation} \noindent
As a consequence, in the context of our GBR method, Athy's law may straightforwardly be reformulated as:
    \begin{equation} \label{GBR::Compaction-AthyGBR}
        \overline\Psi(\overline{\r}_\tau) \ \simeq \ \overline\Psi_o(\overline{\r}_\tau) \cdot exp\biggl\{ \overline\kappa(\overline{\r}_\tau)\cdot{t}_\tau(\overline{\r}_\tau) \biggr\}
        \qquad \forall\ \overline{\r}_\tau\in \overline{G}_\tau
    \end{equation}\noindent

\begin{figure}
\centerline{\psfig{figure=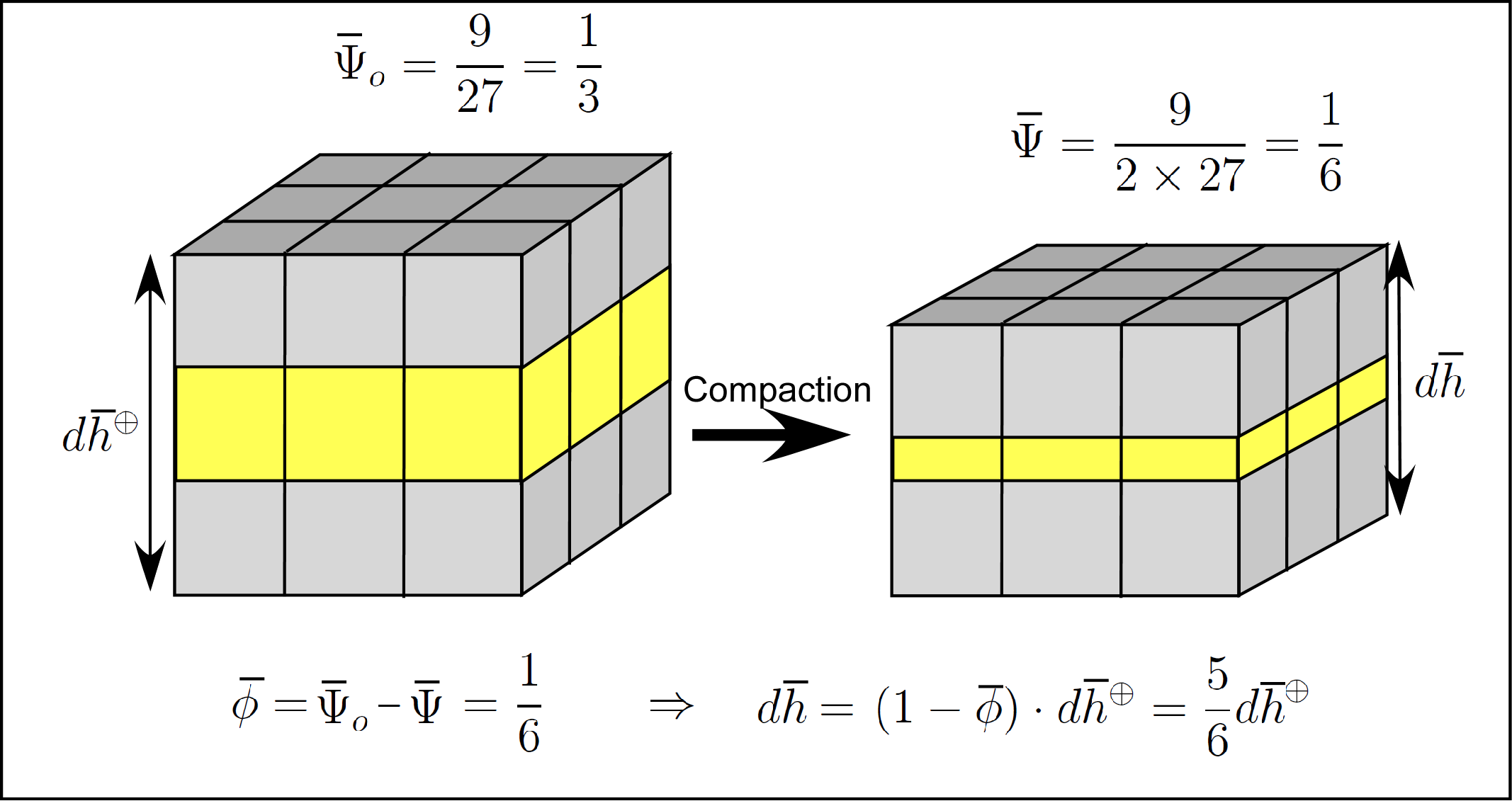,width=116mm}}
\caption{
        Porosity versus compaction: Infinitely small column of
            sediment where, under vertical compaction, the
            vertical dimension of pores represented in yellow is halved. As a
            result, $(d\bar{h}^\oplus=3)$ becomes $(d\bar{h}=2.5)$. Note that
            yellow and gray cells may be randomly swapped in the vertical
            direction without altering these results.
        }
\label{GBR-Compaction}
\end{figure}\noindent

\subsection*{Decompaction in $\Bmath{\overline{G}_\tau}$}

Elasto-plastic mechanical frameworks developed to model compaction rely on a
number of input parameters which may be difficult for a geologist or geomodeler
to assess and require solving a complex system of equations
\citep{Schneider1996}. Isostasic approaches are simpler to parameterize and
still provide useful information on basin evolution \citep{DurandRiard2011}.
Therefore, we will consider compaction as a mainly one-dimensional, vertical
process induced by gravity which mainly occurs in the early stages of sediment burial
when horizons are still roughly horizontal surfaces close to the sea floor. At
any point $\overline{\r}_\tau\in\overline{G}_\tau$ within a layer, the decompacted thickness
$d\bar{h}^\oplus(\overline{\r}_\tau)$ of a vertical probe consisting of an infinitely short column of sediment roughly
orthogonal to the restored horizon passing through
$\overline{\r}_\tau$ is linked to the thickness
$d\bar{h}(\overline{\r}_\tau)$ of the shorter, compacted vertical column by the following
relationship:
\begin{equation} \label{GBR:decompaction-1.u.1}
\forall\ \overline{\r}_\tau\in\overline{G}_\tau \ : \quad
\left|
\begin{array}{c}
    d\bar{h}^\oplus(\overline{\r}_\tau)
    \ = \ \displaystyle
    \frac{1}{1-\overline{\phi}_\tau(\overline{\r}_\tau)}\cdot
    d\bar{h}(\overline{\r}_\tau)
    \\ \\
    \mbox{with : } \overline{\phi}_\tau(\overline{\r}_\tau) = \overline\Psi_o(\overline{\r}_\tau)-\overline\Psi(\overline{\r}_\tau) \in [0,1[
\end{array}
\right.
\end{equation}
\noindent
In this equation, $\overline{\phi}(\overline{\r}_\tau)$ denotes the ``compaction coefficient''
which characterizes vertical shortening of the probe at restored location
$\overline{\r}_\tau\in\overline{G}_\tau$. As an example,
Figure~\ref{GBR-Compaction} shows the same infinitely short vertical column of
    sediment where average porosity is equal to $(\overline\Psi_o=1/3)$ before
    compaction and $(\overline\Psi=1/6)$ after compaction. The compaction
    coefficient $(\overline\Psi_o-\overline\Psi)$ is then equal to
    $(\overline{\phi}=1/6)$ and column shortening $(1-\overline{\phi})$ is
    $(5/6)$.

\subsection*{Taking present day compaction in $\Bmath{\overline{G}_\tau}$ into account}

So far, in the context of our GBR method, the restored $\overline{G}_\tau$-space
has been built assuming that there is no compaction. As a consequence,
$\overline{G}_\tau$ we obtained so far is incorrect because it has undergone
compaction characterized by present day porosity
$\overline{\Psi}_p(\overline{\r}_\tau)$.

\ \\
Let $\bar{\phi}_\tau^\ominus(\overline{\r}_\tau)$ and
$\bar{\phi}_\tau^\oplus(\overline{\r}_\tau)$ be the pair of given functions
defined by:
        \begin{equation} \label{GBR:decompaction-Step-1}
        \forall\ \overline{\r}_\tau\in\overline{G}_\tau \ : \quad
        \left|
        \begin{array}{llllllllllll}
            \bar{\phi}_\tau^\ominus(\overline{\r}_\tau)
            &=&  \bar\Psi_o(\overline{\r}_\tau)-\overline\Psi_p(\overline{\r}_\tau)
            \\ \\
            \bar{\phi}_\tau^\oplus(\overline{\r}_\tau)
             &=& \bar\Psi_o(\overline{\r}_\tau)-\overline\Psi(\overline{\r}_\tau)
        \end{array}
        \right.
        \end{equation}\noindent
where, for coherency with Athy's law, present day porosity $\overline\Psi_p(\overline{\r}_\tau)$ is assumed to honor the following constraint:
        \begin{equation} \label{GBR:decompaction-Step-1.C}
            \overline\Psi_\p(\overline{\r}_\tau) \ \leq \ \overline\Psi(\overline{\r}_\tau)
            \qquad
            \forall\ \overline{\r}_\tau\in\overline{G}_\tau
        \end{equation}\noindent
Note that such a constraint implies that $\phi_\tau^\oplus({\r}_\tau)\leq \phi_\tau^\ominus({\r}_\tau)$.

\ \\
Considering once again the vertical probe introduced above in restored space
$\overline{G}_\tau$ and using equation \ref{GBR:decompaction-1.u.1} twice in a
forward then backward way, to take compaction into account, we propose the
following two steps:
\begin{enumerate}

\item First, to cancel out the compaction characterized by given, present day
porosity $\overline\Psi_p(\overline{\r}_\tau)$, a ``total'', vertical decompaction
is applied by updating $d\bar{h}(\overline{\r}_\tau)$ as follows:
        \begin{equation} \label{GBR:decompaction-Step-1z}
            d\bar{h}_o(\overline{\r}_\tau)
            \ = \ \displaystyle
            \frac{1}{1-\bar{\phi}_\tau^\ominus(\overline{\r}_\tau)}\cdot
            d\bar{h}(\overline{\r}_\tau)
        \end{equation}\noindent
    After this first operation, probe porosity is equal to $\overline{\Psi}_o(\overline{\r}_\tau)$.

\item Next, a ``partial'' recompaction is applied as a function of the actual porosity $\overline\Psi(\overline{\r}_\tau)$
      approximated by Athy's law \ref{GBR::Compaction-AthyGBR} at geological-time $\tau$:
        \begin{equation} \label{GBR:decompaction-Step-1zz}
            d\bar{h}^\oplus(\overline{\r}_\tau)
            \ = \ \displaystyle
            \{1-\bar{\phi}_\tau^\oplus(\overline{\r}_\tau)\}\cdot
            d\bar{h}_o(\overline{\r}_\tau)
        \end{equation}\noindent
    After this second operation, probe porosity is equal to $\overline{\Psi}(\overline{\r}_\tau)$.

\end{enumerate}\noindent
Therefore, to take present-day compaction into account, Equation~\ref{GBR:decompaction-1.u.1} must be replaced by:
\begin{equation} \label{GBR:decompaction-Step-3}
    d\bar{h}^\oplus(\overline{\r}_\tau)
    \ = \ \displaystyle
    \frac{1-\bar{\phi}_\tau^\oplus(\overline{\r}_\tau)}{1-\bar{\phi}_\tau^\ominus(\overline{\r}_\tau)}\cdot
    d\bar{h}(\overline{\r}_\tau)
    \qquad \forall\ \overline{\r}_\tau\in\overline{G}_\tau
\end{equation}
\noindent

\subsection*{GBR approach to decompaction in $\Bmath{\overline{G}_\tau}$}

In the restored $\overline{G}_\tau$-space, ${t}_\tau(\overline{\r}_\tau)$ may be
interpreted as an arc-length abscissa ${s}(\overline{\r}_\tau)$ along the
vertical straight line passing through $\overline{\r}_\tau$ oriented in the same
direction as the vertical unit frame vector\footnote{ See equation \ref{GBR:X00}.}
$\{\overline{\r}_{t_\tau}=\r_z\}$. Therefore, in the $\overline{G}_\tau$-space,
\begin{equation} \label{GBR::Compaction-dt}
 d{t}_\tau(\overline{\r}_\tau)
 \ = \
 ds(\overline{\r}_\tau)
 \ = \
 d\overline{h}(\overline{\r}_\tau)
\end{equation}\noindent
is the height of an infinitely short vertical column of restored sediment
located at point $\overline{\r}_\tau\in\overline{G}_\tau$, subject to present-day compaction.
As a consequence, to take compaction into account in the restored
$\overline{G}_\tau$-space, according to Equations~\ref{GBR:decompaction-Step-3}
and \ref{GBR::Compaction-dt}, function ${t}_\tau(\overline{\r}_\tau)$ must be
replaced by a ``decompacted'' function ${t}_\tau^\oplus(\overline{\r}_\tau)$ such that:
    \begin{equation} \label{GBR:decompaction-A}
          \frac{d{t}_\tau^\oplus}{d{t}_\tau}\bigg|_{\overline{\r}_\tau}
          \ = \ \displaystyle
          \frac{d\bar{h}^\oplus(\overline{\r}_\tau)}{d\bar{h}(\overline{\r}_\tau)}
          \ = \ \displaystyle
          \frac{1-\bar{\phi}_\tau^\oplus(\overline{\r}_\tau)}{1-\bar{\phi}_\tau^\ominus(\overline{\r}_\tau)}
    \end{equation} \noindent
Assuming that $\{\overline{\r}_{t_\tau}=\r_z\}$ is the unit vertical frame vector of the
$\overline{G}_\tau$-space, it is well known that
    \begin{equation} \label{GBR:decompaction-B}
          \grad\,{t}_\tau^\oplus(\overline{\r}_\tau)\cdot \ \overline{\r}_{t_\tau}
          \ = \
          \frac{d{t}_\tau^\oplus(\overline{\r}_\tau + {s}\cdot\overline{\r}_{t_\tau})}{d{s}}\bigg|_{s=0}
          \ = \
           \frac{d{t}_\tau^\oplus}{d{t}_\tau}\bigg|_{\overline{\r}_\tau}
    \end{equation} \noindent
from which we can conclude that the current altitude ${t}_\tau(\overline{\r}_\tau)$
of point $\overline{\r}_\tau\in\overline{G}_\tau$ should be transformed into a
decompacted altitude ${t}_\tau^\oplus(\overline{\r}_\tau)$ honoring the following differential equation:
\begin{equation} \label{GBR:decompaction-C}
\begin{array}{|c|} \hline \\
\quad
          \displaystyle
          \grad\,{t}_\tau^\oplus(\overline{\r}_\tau)\cdot \overline{\r}_{{t}_\tau}
          \ = \ \frac{1-{\phi}_\tau^\oplus(\overline{\r}_\tau)}{1-{\phi}_\tau^\ominus(\overline{\r}_\tau)}
          \qquad \forall\ \overline{\r}_\tau\in \overline{G}_\tau
    \\ \\
    \mbox{with : }\quad
    \bar{\phi}_\tau^\oplus(\overline{\r}_\tau) = \overline{\Psi}_o(\overline{\r}_\tau)-\overline{\Psi}(\overline{\r}_\tau)
    \quad \& \quad
    \bar{\phi}_\tau^\ominus(\overline{\r}_\tau) = \overline{\Psi}_o(\overline{\r}_\tau)-\overline{\Psi}_p(\overline{\r}_\tau)
\quad
\\ \\ \hline \end{array}
\end{equation}
\noindent
Due to the vertical nature of compaction, on $\{\overline{\cal
    S}(0)\equiv\overline{H}_\tau\}$,
function ${t}_\tau^\oplus(\overline{\r}_\tau)$ should vanish and its gradient
should be vertical. In other words, in addition to constraint
\ref{GBR:decompaction-C}, function ${t}_\tau^\oplus(\overline{\r}_\tau)$ must
also honor the following boundary conditions where $\overline{\r}_{u_\tau}$ and
$\overline{\r}_{v_\tau}$ are the unit horizontal frame vectors of the $\overline{G}_\tau$-space:
\begin{equation} \label{GBR:decompaction-D}
\begin{array}{|c|} \hline \\
\quad
\forall\ \overline{\r}_\tau^o\in \{\overline{\cal S}_\tau(0)\equiv \overline{H}_\tau\}\ : \quad
    \left|
    \begin{array}{cc}
          1)&  t_\tau^\oplus(\bar\r_\tau^o) = \ 0
    \\ \\
          2)&  \grad\,{t}_\tau^\oplus(\overline{\r}_\tau^o)\cdot \overline{\r}_{\bar{u}_\tau} \ = \ 0
    \\
          3)&  \grad\,{t}_\tau^\oplus(\overline{\r}_\tau^o)\cdot
          \overline{\r}_{\bar{v}_\tau} \ = \ 0
    \end{array}
    \right.
\quad
\\ \\ \hline \end{array}
\end{equation}
\noindent
As compaction is a continuous process, ${t}_\tau^\oplus(\overline{\r}_\tau)$
must be ${\cal C}^o$-continuous across all faults affecting $\overline{G}_\tau$.
As a consequence, in addition to constraints \ref{GBR:decompaction-C} and
\ref{GBR:decompaction-D}, for any fault $\overline{F}$ in $\overline{G}_\tau$, function
${t}_\tau^\oplus(\overline{\r}_\tau)$ must also honor the following boundary
conditions where $(\overline{\r}_{\msc{f}}^\oplus,\overline{\r}_{\msc{f}}^\ominus)_\tau$
are pairs of ``$\tau$-mate-points''defined as collocated points lying on the
positive face $\overline{F}^+$
and negative face $\overline{F}^-$ of $\overline{F}$ at geological-time $\tau$:
            \begin{equation} \label{GBR:decompaction-E}
            \begin{array}{|c|} \hline \\
                  \begin{array}{c}
                    {t}_\tau(\overline{\r}_{\msc{f}}^\oplus) \ = \ {t}_\tau(\overline{\r}_{\msc{f}}^\ominus)
                    \\ \\
                    \forall\ \overline{F}\in \overline{G}_\tau \quad \&\quad
                    \forall\
                        (\overline{\r}_{\msc{f}}^\oplus,\overline{\r}_{\msc{f}}^\ominus)_\tau \in \ \overline{F}
                  \end{array}
            \\ \\ \hline \end{array}
            \end{equation}
            \noindent
Using an appropriate numerical method, ${t}_\tau^\oplus(\r_\tau)$ must be
computed in $\overline{G}_\tau$ whilst ensuring that differential
equation~\ref{GBR:decompaction-C} and boundary conditions
\ref{GBR:decompaction-D} and \ref{GBR:decompaction-E} are honored. To ensure
smoothness and uniqueness of
${t}_\tau^\oplus(\r_\tau)$, the following constraint may also be added:
                \begin{equation} \label{GBR:decompaction-F}
                                \sum_{(a,b)\in \{u_\tau,v_\tau, t_\tau\}^2}
                                \int_{\overline{G}_\tau}
                                    \biggl\{\partial_a\partial_b\, {t}_\tau^\oplus(\overline{\r}_\tau) \biggr\}^2
                                    \cdot d\overline{\r}_\tau
                                \qquad \mbox{minimum}
                \end{equation}\noindent

\ \\
As a conclusion, to take compaction into account, the following GBR approach may be used:
\begin{enumerate}

\item Compute a numerical approximation of ${t}_\tau^\oplus(\overline{\r}_\tau)$ in $\overline{G}_\tau$ and use the reverse $u_\tau v_\tau t_\tau$-transform to update ${t}_\tau({\r}_\tau)$ in ${G}_\tau$:
                \begin{equation} \label{GBR:decompaction-G}
                      {t}_\tau({\r}_\tau) \longleftarrow \ {t}_\tau^\oplus(\overline{\r}_\tau)
                      \qquad \forall\ {\r}_\tau\in G_\tau;
                \end{equation}\noindent

\item Recompute numerical approximations of restoration functions $u_\tau(\r_\tau)$ and $v_\tau(\r_\tau)$ in $G_\tau$ to prevent voids
and overlaps in the restored space, as,
according to Equations~\ref{GBR:X11} and \ref{GBR:S1}, $u_\tau(\r_\tau)$ and $v_\tau(\r_\tau)$
depend on $t_\tau(\r_\tau)$;

\item Build the ``decompacted'' restored space $\overline{G}_\tau$ as the new, direct $u_\tau v_\tau t_\tau$-transform of geological space $G_\tau$ observed today.

\end{enumerate}\noindent

\ \\
This approach to decompaction is fully derived from the GBR framework described
in this paper and differs from the sequential decompaction following
Athy's law along IPG-lines applied by~\citet{Lovely2018}.

\section{Constraints summary}

Among all the equations presented so far,
Equations~\ref{GBR:eikonal-equation}-1, \ref{GBR:Indet}, \ref{GBR:Indet-1} and \ref{GBR-Sigma-6}
are the most critical.

\ \\
First and above all, honoring constraint~\ref{GBR:eikonal-equation}-1 as
closely as possible is the very heart of the proposed {GBR} method. Due to local
deformations of horizons, this equation may generally be honored only in a
least squares sense.
However, if $||\grad\,t_\tau||_\r$ deviates too much from 1, then, during the
restoration process, layer thicknesses will not be preserved, which may induce
undesirable volume variations; and due to constraints~\ref{GBR:X11} or
\ref{GBR:S1} based on $t_\tau(\r_\tau)$, restoration functions $\{u_\tau,v_\tau\}_{\r_\tau}$ will be
incorrect.

\ \\
Next, constraints~\ref{GBR-Sigma-6} are of paramount importance because,
during restoration of horizon $H_\tau$, they prevent gaps and overlaps from
appearing in $\overline{G}_\tau$ along faults.

\ \\
Finally, constraints~\ref{GBR:Indet} and \ref{GBR:Indet-1} are also extremely important because
they preserve coherency of restored surface $\overline{H}_\tau$ viewed either as
the $u_\tau v_\tau t_\tau$-transform or the regular GeoChron $uvt$-transform of
$H_\tau$. Without constraints~\ref{GBR:Indet}, the GBR method would not be
consistent with the input GeoChron model.

\subsection*{Comment: Volume preservation}

Barring the effects of compaction, let us consider, in the $G_\tau$-space, a
pseudo-layer $L(d,\varepsilon)$ with infinitely small thickness $\varepsilon$
bounded by pseudo-horizons ${\cal S}(d)$ and ${\cal S}(d-\varepsilon)$. Because
of eikonal constraint \ref{GBR:eikonal-equation}, in the
$\overline{G}_\tau$-space, restored layer $\overline{L}(d,\varepsilon)$ holds
as closely as possible the same thickness $\varepsilon$ as $L(d,\varepsilon)$.

\ \\
Consider now, in the $G_\tau$-space, an infinitely small compact patch
$\Delta{\cal S}(d)$ drawn on ${\cal S}(d)$ and let $\Delta{\cal
S}(d-\varepsilon)$ be the projection of this patch onto ${\cal
S}(d-\varepsilon)$ along lines with constant
$\{u_\tau,v_\tau\}$ coordinates\footnote{ In GeoChron theory, these lines are
called ``Iso-Paleo-Geographic'' lines and abbreviated IPG-lines.}
passing through ${\cal S}(d)$. Let $\Delta V(d,\varepsilon)$ be the infinitely
small volume bounded by $\Delta{\cal S}(d)$, $\Delta{\cal S}(d-\varepsilon)$ and
the field of lines defined above. During restoration, depending on the
structural style, two cases have to be considered:
\begin{itemize}

\item if the structural style is flexural slip, by definition\footnote{ See
\citet{Mallet2014}, page 72.}, areas and angles
on surfaces ${\cal S}(d)$ and ${\cal S}(d-\varepsilon)$ are preserved;

\item if the structural style is minimal deformation, by definition\footnote{ See
\citet{Mallet2014}, page 71.}, deformations
of areas and angles on surfaces ${\cal S}(d)$ and ${\cal S}(d-\varepsilon)$ are
minimized, as much as possible.

\end{itemize}\noindent
Therefore, omitting compaction, as in both cases thickness $\varepsilon$ is
preserved as much as possible, volumes of $\Delta V(d,\varepsilon)$ and its restored version
$\overline{\Delta} V(d,\varepsilon)$ are as identical as possible.

\section{Numerically approximating $\protect\Bmath{\{u_\tau,v_\tau,t_\tau\}}$}

\label{GBR-Improving-t-tau}

From a theoretical standpoint, restoration functions
$\{u_\tau,v_\tau,t_\tau\}_{\r_\tau}$ are solutions to a wide system of partial
differential equations presented so far in this paper. However, from a practical
perspective, these equations are often non linear and coupled, which makes them
difficult to solve. Many general numerical techniques known in the art
could be employed but, as we show in the following, the geological nature of our
problem makes it possible for us to replace these complex
differential equations by surrogates which are easier to solve.

\subsection*{About the eikonal equation}

\begin{figure}
\centerline{\psfig{figure=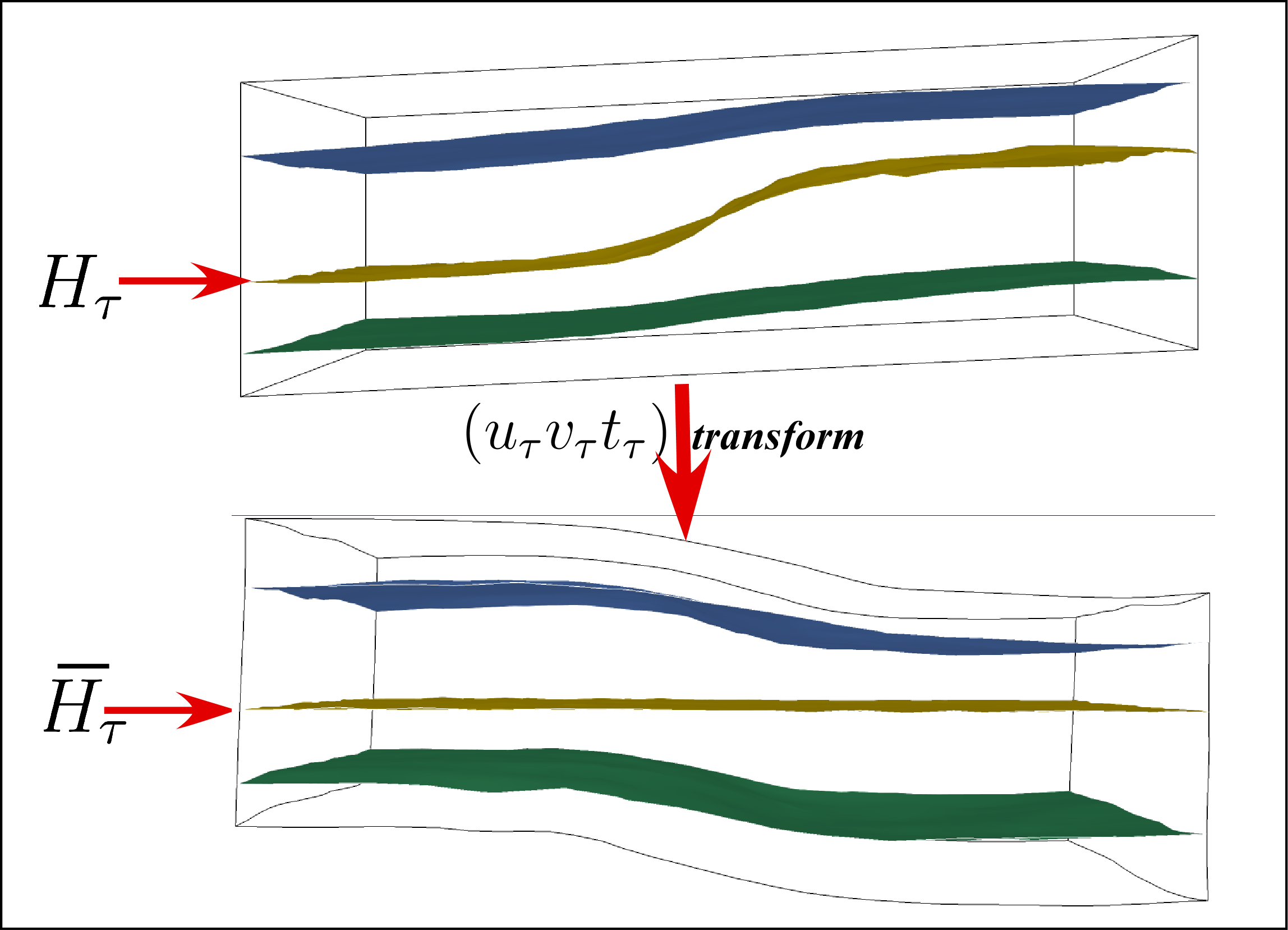,width=116mm}}
\caption{
        Test example: {GBR} of a ``ramp'' structure. Restored horizon $H_\tau$ is the central, sigmoid surface. Note that main curvature of $H_\tau$ locally varies and the associated curvature center moves from one side to the opposite side of $H_\tau$.
        }
\label{GBR-Ramp-test}

\end{figure}

As pointed out in the previous section, computing a function $t_\tau(\r_\tau)$ which honors eikonal
Equation~\ref{GBR:eikonal-equation} is the corner-stone of our
proposed {GBR} method but Equation~\ref{GBR:eikonal-equation}-1, recalled below,
is not linear:
    \begin{equation} \label{GBR:eikonal-equation-1}
        ||\grad\,t_\tau(\r_\tau)|| = 1 \qquad \forall\ \r_\tau\in G_\tau
    \end{equation}\noindent
Through Equations~\ref{GBR:X11} or \ref{GBR:S1}, any excessive violation of this
constraint also impacts functions $\{u_\tau,v_\tau\}_{\r_\tau}$ and the resulting
restoration is then inevitably incorrect.

\ \\
Based on the test example shown in Figure~\ref{GBR-Ramp-test}, where
horizon $H_\tau$ to restore is the central sigmoid surface, results obtained
with two different numerical techniques are compared and shown on
Figure~\ref{GBR-Ramp-results}.
This seemingly simple test is actually highly significant because it shows
local variations in curvature which make eikonal Equation~\ref{GBR:eikonal-equation-1} difficult to approximate numerically.

\noindent

\subsection*{Computing $\Bmath{t_\tau(\r_\tau)}$: Surrogate (weak) eikonal equations}

We have stated before that:
    \begin{equation} \label{GBR:X8}
            \grad\,t_\tau(\r_\tau^o) \ = \ \B{N}(\r_\tau^o) \ = \ \frac{\grad\,t(\r_\tau^o)}{||\grad\,t(\r_\tau^o)||}
            \qquad \forall\ \r_\tau^o\in H_\tau
    \end{equation}\noindent
which means that eikonal Equation~\ref{GBR:eikonal-equation} is approximately
equivalent to the following system called ``surrogate-eikonal'' equation:
        \begin{equation} \label{GBR:X9}
        \left|
              \begin{array}{cccc}
                    1)& \displaystyle
                                \sum_{\alpha\in\{x,y,z\}}
                                \int_{G_\tau} ||\partial_\alpha\, \grad\, t_\tau(\r_\tau) ||^2  \cdot d\r_\tau \qquad \mbox{minimum}
                    \\ \\
                    2)& \mbox{subject to : \ }
                        \left\{
                        \begin{array}{cclllll}
                            a)&   t_\tau(\r_\tau^o) = 0
                            \\ \\
                            b)&  \displaystyle \grad\,t_\tau(\r_\tau^o) = \B{N}(\r_\tau^o)
                        \end{array}
                        \right\} \quad \forall\ \r_\tau^o\in  H_\tau
              \end{array}\noindent
        \right.
        \end{equation}\noindent
Eikonal Equations~\ref{GBR:eikonal-equation}-2 are strictly honored
on $H_\tau$ and Equation~\ref{GBR:X9}-1 is assumed to smoothly
propagate $\grad\,t_\tau(\r)$ in such a way that, everywhere inside $G_\tau$ and
similarly to Equation~\ref{GBR:eikonal-equation}-1, $\grad\,t_\tau(\r_\tau)$
roughly remains a unit vector field.
In practice, according to techniques known in the art, Equation~\ref{GBR:X9}-1 may be
linearly approximated so that each Equation~\ref{GBR:X9} is linear and,
therefore, easier to solve than ``true'' eikonal equation ~\ref{GBR:eikonal-equation}.

\ \\
As mentioned above, at any point $\r_\tau\in G_\tau$, Equation~\ref{GBR:X9}-1 should
ensure that $||\grad\,t_\tau(\r_\tau)||$ is equal to its unit starting value on
$H_\tau$. Unfortunately, away from $H_\tau$, numerical drift usually makes
$||\grad\,t_\tau(\r_\tau)||$ deviate from target value 1. As a consequence, eikonal
constraint~\ref{GBR:eikonal-equation}-1 is generally not perfectly honored away from
$\{{\cal S}_\tau(0)\equiv H_\tau\}$, which implies that, after restoration,
distortions inevitably appear in the vertical direction of the
$\overline{G}_\tau$-space.

\begin{figure}
\centerline{\psfig{figure=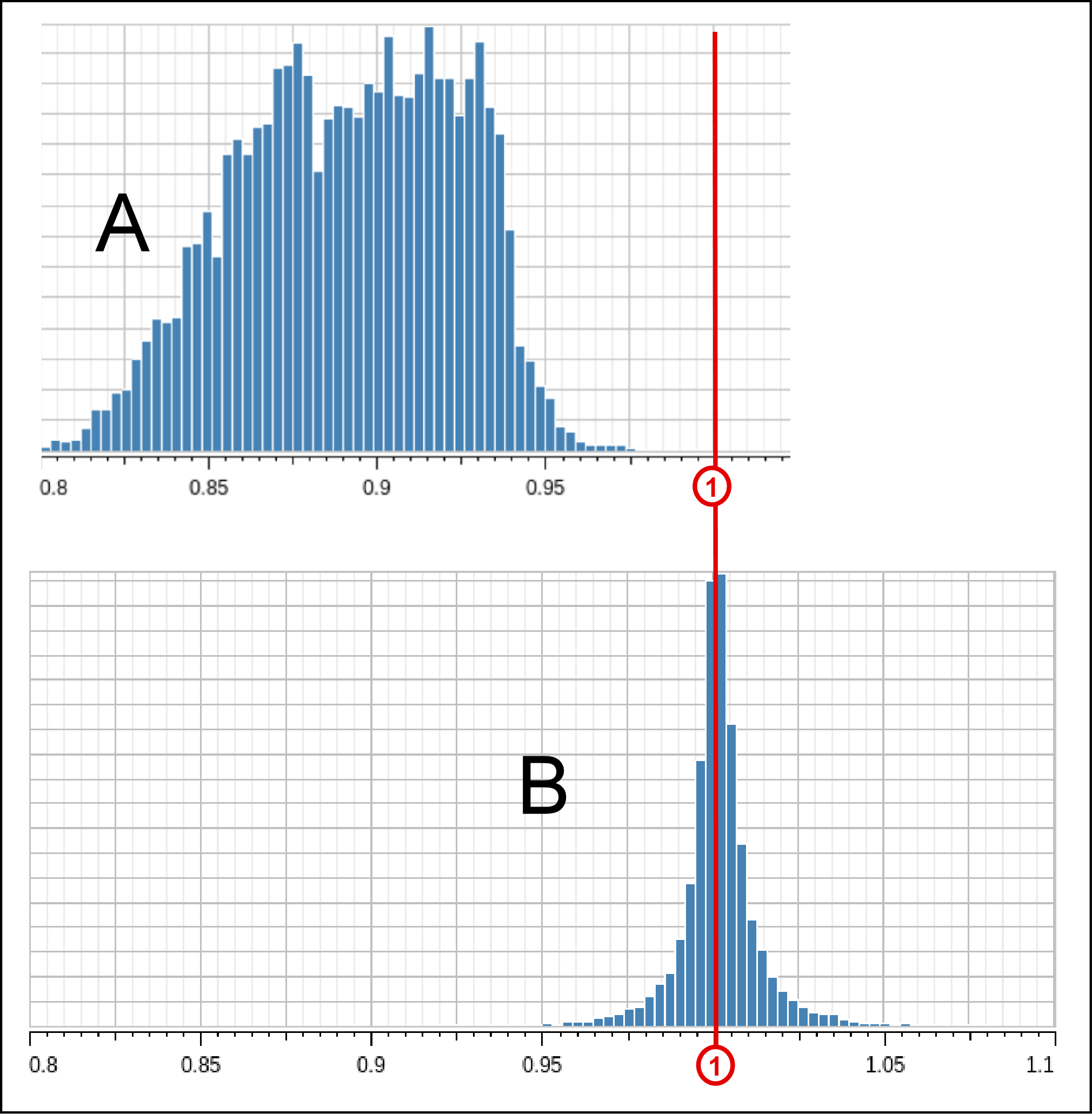,width=100mm}}
\caption{
        Test example showing histograms of $||\grad\,t_\tau(\r_\tau)||$ in the
studied domain corresponding to Figure~\ref{GBR-Ramp-test}. Depending on the
method used to compute $t_\tau(\r_\tau)$, the resulting magnitude of
$||\grad\,t_\tau(\r_\tau)||$ may severely deviate from value ``1'' required by
eikonal Equation~\ref{GBR:eikonal-equation}-1. Note that, for clarity's sake,
the vertical axis of histogram (B) has been shrunk by a factor of 3.
        }
\label{GBR-Ramp-results}

\end{figure}\noindent

\ \\
On Figure~\ref{GBR-Ramp-results}-A, the histogram of $||\grad\,t_\tau(\r_\tau)||$
so obtained in $G_\tau$ with surrogate eikonal Equations~\ref{GBR:X9} applied to our test example clearly shows that
eikonal Equation~\ref{GBR:eikonal-equation}-1 is not honored correctly.
First, $||\grad\,t_\tau(\r_\tau)||$ is never equal to 1. Next, the median value is about
0.89, which represents an error of 11~\%. Finally, standard deviation is 0.032
and the spread between 25$^{th}$ and 75$^{th}$ percentiles is 0.051.

\ \\
We can conclude from these figures that approximating eikonal
Equation~\ref{GBR:eikonal-equation}-1 by Equations~\ref{GBR:X9} does not give
precise enough results.

\subsection*{Computing $\Bmath{t_\tau(\r_\tau)}$: A precise incremental solution}

Generally, even though eikonal Equation~\ref{GBR:eikonal-equation}-1 is not
perfectly honored, function $t_\tau(\r_\tau)$ generated by
Equations~\ref{GBR:X9} may be considered as an approximation of
the actual solution. In other words, assuming that $t_\tau^\star(\r)$ is a first approximation of $t_\tau(\r_\tau)$, there is an unknown function
$\varepsilon_\tau(\r_\tau)$ which may be used as follows to compute, in a post-processing step, an improved version of $t_\tau(\r_\tau)$:
                  \begin{equation} \label{GBR-ImprovT-10}
                    t_\tau(\r_\tau) \ = \ t_\tau^\star(\r_\tau) +
                    \varepsilon_\tau(\r_\tau)
                  \end{equation}
                  \noindent
with
                  \begin{equation} \label{GBR-ImprovT-11}
                        \varepsilon_\tau(\r_\tau^o) \ = \ - t_\tau^\star(\r_\tau^o)
                             \qquad\ \forall\ \r_\tau^o\in H_\tau
                  \end{equation}
                  \noindent
and where $t_\tau^\star(\r_\tau)$ is assumed to be precise enough to honor:
                  \begin{equation} \label{GBR-ImprovT-11-XX}
                        ||\grad\, \varepsilon_\tau(\r_\tau)|| \ \ll  \ ||\grad\, t_\tau^\star(\r_\tau)|| \ \simeq \ 1  \qquad\ \forall\ \r_\tau\in G_\tau
                  \end{equation}
                  \noindent

\ \\
Through faults, $\varepsilon_\tau(\r_\tau)$ is assumed to behave in a
similar way to function $t_\tau(\r_\tau)$. In other words,
referring to constraints~\ref{GBR-Secondary-faults}, for any pair of
$\tau$-mate-points $(\r_{\msc{f}}^\oplus,\r_{\msc{f}}^\ominus)$ located on a $\tau$-inactive fault $F$,
function $\varepsilon_\tau(\r_\tau)$ and its gradient must honor the following
equations:
            \begin{equation} \label{GBR-Secondary-faults-XXX}
                  \left.
                  \begin{array}{cccccccccccccccccc}
                    1)& \varepsilon_\tau(\r_{\msc{f}}^\oplus) - \varepsilon_\tau(\r_{\msc{f}}^\ominus)
                    &=& t_\tau^\star(\r_{\msc{f}}^\ominus) - t_\tau^\star(\r_{\msc{f}}^\oplus)
                    \\ \\
                    2)& \grad\,\varepsilon_\tau(\r_{\msc{f}}^\oplus) - \grad\,\varepsilon_\tau(\r_{\msc{f}}^\ominus)
                    &=& \grad\,t_\tau^\star(\r_{\msc{f}}^\ominus) - \grad\,t_\tau^\star(\r_{\msc{f}}^\oplus)
                  \end{array}
                  \right\}
                \quad \forall\ (\r_{\msc{f}}^\oplus,\r_{\msc{f}}^\ominus) \in \mbox{$\tau$-inactive fault}
            \end{equation}
            \noindent

\ \\
In addition to constraints~\ref{GBR-ImprovT-11} and
\ref{GBR-Secondary-faults-XXX}, to better fit eikonal
Equation~\ref{GBR:eikonal-equation}-1, the unknown function $\varepsilon_\tau(\r_\tau)$
should also honor the following non linear constraint:
                  \begin{equation} \label{GBR-ImprovT-12}
                    1 \ = \
                    ||\grad\,\{t_\tau^\star(\r_\tau)+\varepsilon_\tau(\r_\tau)\}||^2 \ = \
                    ||\grad\,t_\tau^\star(\r_\tau)||^2 \ + \
                    ||\grad\,\varepsilon_\tau(\r_\tau)||^2 \ + \
                    2\cdot \grad\,t_\tau^\star(\r_\tau) \cdot \grad\,\varepsilon_\tau(\r_\tau)
                  \end{equation}
                  \noindent
According to Equation~\ref{GBR-ImprovT-11-XX}, second order term
$||\grad\,\varepsilon_\tau(\r_\tau)||^2$ may be neglected in order to linearize the
equation above:
                  \begin{equation} \label{GBR-ImprovT-13}
                    \quad
                        \grad\,\varepsilon_\tau(\r_\tau) \cdot \grad\,t_\tau^\star(\r_\tau) \ \simeq \
                        \frac{1}{2}\cdot
                        \{
                            1 - ||\grad\,t_\tau^\star(\r_\tau)||^2
                        \}
                        \qquad \forall\ \r_\tau\in G_\tau
                    \quad
                  \end{equation}
                  \noindent
This linear constraint to be honored in a least squares sense, in addition to
constraints~\ref{GBR-ImprovT-11} and \ref{GBR-Secondary-faults-XXX}, fully
characterizes function $\varepsilon_\tau(\r_\tau)$ in a unique way. Similarly to
Equation~\ref{GBR:X9}-1, the following constraint may be added to ensure
$\varepsilon_\tau(\r_\tau)$ is smooth:
                \begin{equation} \label{GBR:Epsilon-smooth}
                \sum_{\alpha\in\{x,y,z\}}
                            \int_{G_\tau} ||\partial_\alpha\, \varepsilon_\tau({\r}_\tau)||^2  \cdot d\r_\tau
                            \qquad \mbox{minimum}
                \end{equation}\noindent

\ \\
On Figure~\ref{GBR-Ramp-results}-B, the histogram of $||\grad\,t_\tau(\r_\tau)||$
obtained on our test example with the above incremental approach shows
that eikonal Equation~\ref{GBR:eikonal-equation}-1 is now correctly honored:
$||\grad\,t_\tau(\r_\tau)||$ is, in average, very close to 1. The median value stands
at 1.0, standard deviation is 0.012 and the spread between 25$^{th}$ and
75$^{th}$ percentiles is reduced to 0.0092.

\ \\
From these observations, we can conclude that the above incremental
approximation of eikonal Equation~\ref{GBR:eikonal-equation}-1 is well suited
to computing function $t_\tau(\r_\tau)$.
Similar results may be observed on other test examples of varying complexity.

\subsection*{Computing $\protect\Bmath{\{u_\tau, v_\tau\}_{\r_\tau}}$}

Assuming that $t_\tau(\r_\tau)$ has already been numerically approximated, to
compute an approximation of $\{u_\tau,v_\tau\}_{\r_\tau}$, our approach derives from a technique
suggested on page 123 of \citep{Mallet2014}:
\begin{enumerate}

\item assuming that $\B{N}_\tau(\r)$ is defined as follows in $G_\tau$:
                \begin{equation} \label{GBR-AxeCoaxe-2}
                    {\bf N}_\tau(\r_\tau) \ = \
                    \frac{  \grad\,t_\tau(\r_\tau)  }
                         {||\grad\,t_\tau(\r_\tau)||}
                    \qquad \forall\ \r_\tau\in G_\tau
                \end{equation}\noindent
we compute global structural axis $\B{A}_\tau$ defined as a unit vector
averagely orthogonal to vector field $\B{N}_\tau(\r_\tau)$;

\item for any point $\r_\tau\in G_\tau$, we compute local structural axis
$\Bmath{a}_\tau(\r_\tau)$ and co-axis $\Bmath{b}_\tau(\r_\tau)$ as follows:
                \begin{equation} \label{GBR-AxeCoaxe-1}
                    \Bmath{a}_\tau(\r_\tau) \ = \
                    \frac{  {\bf N}_\tau(\r_\tau)\times{\bf A}_\tau\times{\bf N}_\tau(\r_\tau)  }
                         {||{\bf N}_\tau(\r_\tau)\times{\bf A}_\tau\times{\bf N}_\tau(\r_\tau)||}
                         \, ; \qquad
                    \Bmath{b}_\tau(\r_\tau) \ = \ {\bf
                        N}_\tau(\r_\tau)\times\Bmath{a}_\tau(\r_\tau) \, ;
                \end{equation}\noindent

\item depending on tectonic style, for any point $\r_\tau\in G_\tau$, restoration functions
$\{u_\tau,v_\tau\}_{\r_\tau}$ are set to honor the following surrogate equations in a
least squares sense:
    \begin{itemize}

        \item in a minimal deformation context, Equation~\ref{GBR:X11} may be
        approximated by:
                \begin{equation} \label{GBR-MinDefStyle}
                \left|
                \begin{array}{ccccccccccccc}
                    \grad\,u_\tau(\r_\tau) &\times& \Bmath{a}_\tau(\r_\tau)  &\simeq& \B{0}\\
                    \grad\,v_\tau(\r_\tau) &\times& \Bmath{b}_\tau(\r_\tau)
                    &\simeq& \B{0}
                \end{array}
                \right.
                \end{equation}\noindent

        \item in a flexural slip context, Equation~\ref{GBR:S1} may be
        approximated by:
                \begin{equation} \label{GBR-FlexSlipStyle}
                \left|
                \begin{array}{cccccccccccc}
                    \grad_{\textsc{s}}\,u_\tau(\r_\tau) &\times& \Bmath{a}_\tau(\r_\tau) &\simeq& \B{0} \\
                    \grad_{\textsc{s}}\,v_\tau(\r_\tau) &\times&
                    \Bmath{b}_\tau(\r_\tau) &\simeq& \B{0}
                \end{array}
                \right.
                \end{equation}\noindent

    \end{itemize}\noindent
    In the particular case where $H_\tau$ is a perfect cylindrical surface, it can be shown that these approximations are exact.
        Compared to similar Equations~3.124 and 3.125 on page 123 of \citep{Mallet2014},
        the surrogate equations above have been slightly adapted not to conflict
        with Equation~\ref{GBR:Indet-1} on $H_\tau$;

\item finally, to ensure smoothness and uniqueness of functions $\{u_\tau,v_\tau\}_{\r_\tau}$,
    constraint~\ref{GBR-UtauVtau-smoothness} is added.

\end{enumerate}\noindent
In practice, numerical results so obtained generally yield sufficiently precise
approximations for restoration functions $\{u_\tau,v_\tau\}_{\r_\tau}$. If more
precision is required, these approximations could be improved with an incremental technique
similar to the one proposed above for restoration function $t_\tau(\r_\tau)$.

\section{Examples of 3D restoration}

\label{GBR-Validating-1}

\begin{figure}
\centerline{\psfig{figure=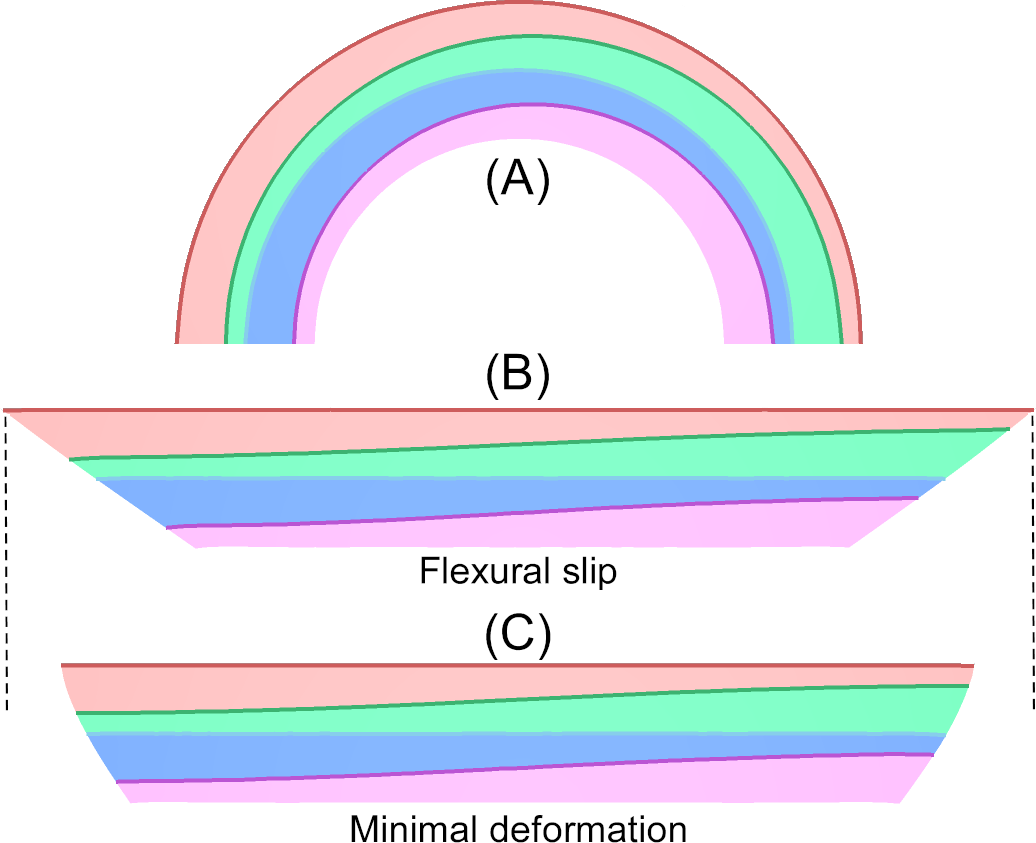,width=100mm}}
\caption{
Vertical cross-section of a three-dimensional synthetic model with four horizons
showing both variations in layer thickness and extreme deformation as towards
the extremities of the model, horizons become vertical (A). Restoration using
the flexural slip tectonic style (B) results in better conservation of horizon
area than using the minimal deformation style (C).
}
\label{rainbow}
\end{figure}

\begin{figure}
\centerline{\psfig{figure=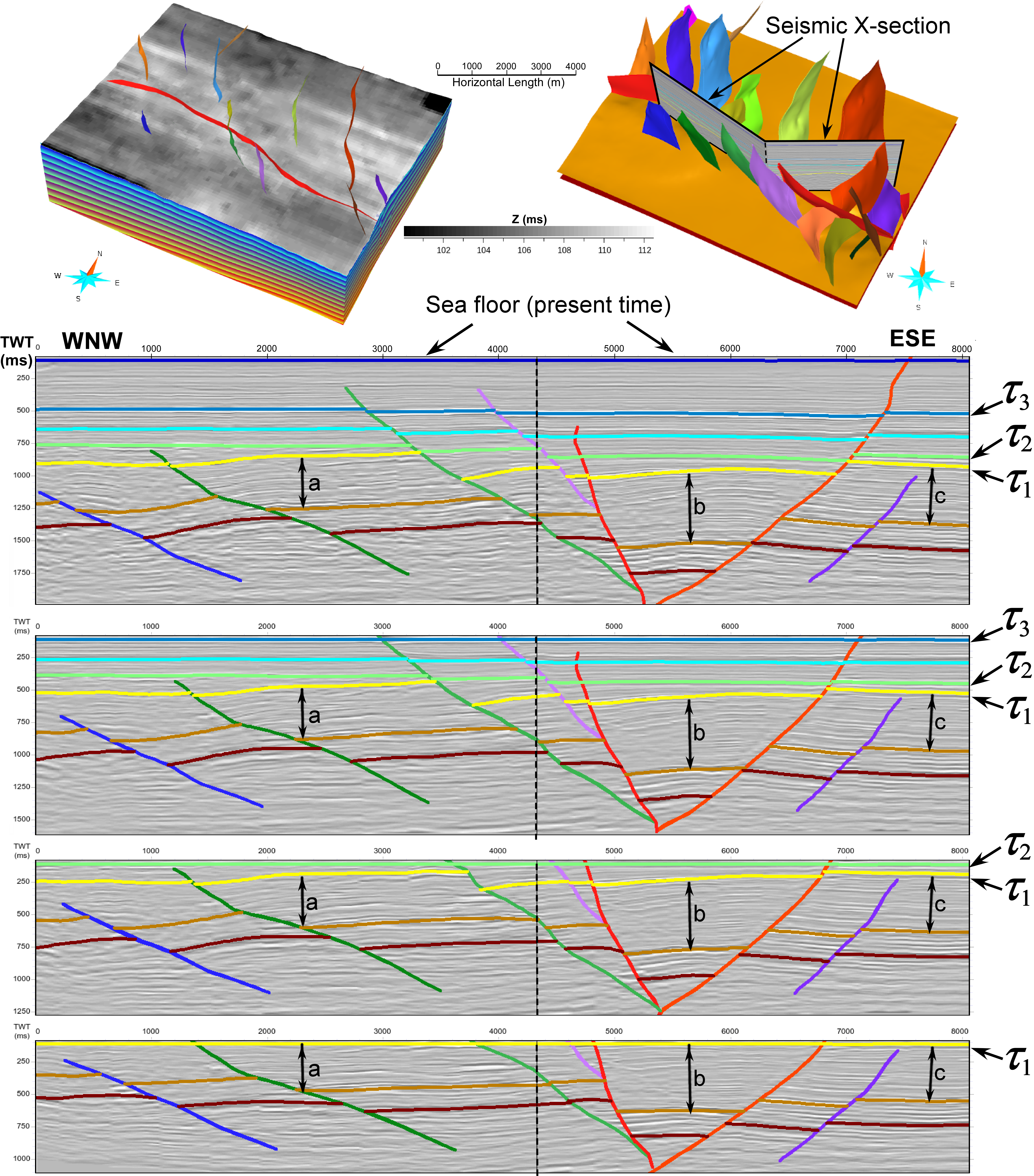,width=170mm}}
\caption{
Restoration of several horizons in a three-dimensional model with an erosive
stratigraphic sequence. Location of seismic image section is shown in top right
corner. Sections below are painted with seismic image, faults and horizons in the
present-day model and at three restoration times. Interval times between two
horizons highlighted by thick, black arrows are preserved.
}
\label{clyde_sections}
\end{figure}

Figure~\ref{rainbow} shows the restoration of a synthetic model with four
horizons, modeled on a grid with about 84,000 cells. Restoration functions
$\{u_\tau,v_\tau,t_\tau\}_{\r_\tau}$ and the associated restoration vector field
$\B{R}_\tau(\r_\tau)$ are computed on the grid for each
restoration time $\tau$ from the initial GeoChron functions $\{u,v,t\}_{\r}$
using the framework and algorithms described in this paper. The full structural
model is then updated on demand to reflect the restored state specified by the
user.
This synthetic example was
designed to illustrate the correct behavior of the GBR method on layers with
varying thickness and horizons with extreme deformation as their extremities on
either side are vertical.

\ \\
Total computation time on an average workstation is 2.25~s per horizon to
restore. Switching between two restored states then takes 0.07~s. Using the
flexural slip tectonic style, variations in area for horizons from
present-day state (A) to restored state (B) are -3.91~\% for the top horizon and
+0.225~\%
for the bottom horizon. As expected, areal variations are higher if the minimal
deformation tectonic style (C) is used (-14.9~\% and +9.10~\% for top and bottom
horizons, respectively). The neutral axis in this model when the minimal
deformation regime is applied is located close to the third, blue horizon at the
top of the blue layer for which areal variation at this restoration stage
is 0.243~\%. As this model is essentially a two-dimensional example
with no variation in geometry in the third dimension, volume variation figures
are similar, with a -1.67~\% global volume variation between initial and restored
states for top horizon in the flexural slip case and -4.46~\% in the minimal
deformation case.

\ \\
This extreme example illustrates that restoration results depend on the initial
GeoChron paleo-coordinates $\{u,v\}_\r$ from which restoration functions
$\{u_\tau,v_\tau\}_{\r_\tau}$ are computed. When the tectonic style is minimal
deformation, specifying the location of the horizon with the minimal amount of
deformation in the initial GeoChron model would help compute $\{u_\tau,v_\tau\}_{\r_\tau}$
such that in restored states, deformations on that specific horizon are
minimized.

\ \\
Figure~\ref{clyde_sections} shows a
full structural volume model restored to deposition time of various horizons.
On an average workstation, computation time in this grid with 845,150
cells was 29.8~s per horizon to restore. Switching from one
restored state to the next then takes 1.15~s.

\ \\
The top-left block diagram shows the model restored to present-day sea floor
geometry, used as an approximation of paleo-topography. The horizon being
restored is an erosive surface and the volume below shows the geological-time
function for the eroded terrains. The image to the right shows the location of a
seismic cross section rendered at different restored times $\{\tau_3, \tau_2,
\tau_1\}$. The top cross section is the present-day geometry of horizons and
faults painted over the seismic image. The cross sections below show horizons,
faults and seismic image restored at times $\tau_3$ when the blue horizon was
deposited, $\tau_2$ when the green, erosive horizon was deposited and $\tau_1$
when the yellow, first horizon modeled in the eroded sequence was deposited.

\ \\
Each restored model is consistent: Despite the complexity of the fault network,
there are no gaps between faults and horizons and no overlaps between fault
blocks. Interval times between horizons, highlighted by identical black arrows
on each cross section, are a constant 360~ms for \textbf{a}, 500 ms for
\textbf{b} and 395~ms for \textbf{c}.

\section{Conclusions}

In this paper, we propose a new restoration method based on the GeoChron model.
Contrary to classical, mechanical methods based on elasticity theory,
this new method is purely geometrical and, therefore, does
not require prior knowledge of geo-mechanical properties of the terrains. This
method works equally well for small and large deformations and for any possible
mechanical behavior (elastic, plastic, \ldots) of the terrains. Moreover, the
restoration process in itself handles consistency around faults and with the
tectonic style chosen by the geomodeler. Finally, a new technique aimed at
taking compaction into account is also proposed.

\ \\
This restoration method also requires less
computation and fewer user inputs than classical geo-mechanical methods. As a
consequence, it is fast and simple to use, which allows
geologists to routinely check and validate structural model
consistency. At any given geological-time $\tau$, if inconsistencies are
spotted, the geological-time function $t(\r)$ ruling the geometry of the
horizons of a restored GeoChron model may be locally interactively edited. Such
changes of $t(\r)$ can automatically and instantly
be back-propagated to the initial GeoChron model corresponding to the
present-day subsurface, without any additional computations.

\section{Acknowledgments}
The authors would like to thank Emerson for their support and for permission to
publish this paper.

\noindent

\noindent


\bibliographystyle{aapg}
\bibliography{restoration}

\end{document}